\documentclass[twocol]{ametsocV6.1}

\usepackage{hyperref}
\hypersetup{
	breaklinks,
    colorlinks=true,
	allcolors=black,
 }

\ifdraft
\else
\usepackage[usenames,dvipsnames]{xcolor}
\PassOptionsToPackage{hyphens}{url}
\hypersetup{
	breaklinks,
	colorlinks=true,
	linkcolor=blue,
    urlcolor=Purple,
	citecolor=Purple,
 }
\fi

\usepackage{siunitx, doi}

\renewcommand{\equiv}{\ensuremath{\stackrel{\mathrm{def}}{=}}} 

\title{Evaluating and improving wave and non-wave stress parametrisations for oceanic flows}

\authors{Daniel R. Johnston,\aff{a,b}\correspondingauthor{Daniel R. Johnston, daniel.johnston@unsw.edu.au}
Callum J. Shakespeare,\aff{b} and
Navid C. Constantinou,\aff{c,d}
}

\affiliation{\aff{a}{School of Science, UNSW Canberra, Australia}\\
\aff{b}{Research School of Earth Sciences, Australian National University, Australia}\\
\aff{c}{School of Geography, Earth and Atmospheric Sciences, University of Melbourne, Australia}\\
\aff{d}{Australian Research Council Centre of Excellence for the Weather of the 21st Century, University of Melbourne, Australia}
}

\abstract{Whenever oceanic currents flow over rough topography, there is an associated stress that acts to modify the flow. In the deep ocean, this stress is predominantly a form drag due to pressure differentials across topography, caused by the formation of internal waves and other baroclinic motions: processes that act on such small scales most global ocean models cannot resolve. Despite the need to incorporate this stress into ocean models, existing parametrisations are limited in their applicability. For instance, most parametrisations are only suitable for small-scale topography and are either for periodic or steady flows, but rarely a combination thereof. Here we summarise some of the most widely used parametrisations and evaluate the accuracy of a carefully selected subset using hundreds of idealised two-dimensional and three-dimensional simulations spanning a wide parameter space. We focus on the case of an isolated Gaussian hill as an idealised representation of a seamount. In cases where the parametrisations prove to be inaccurate, we use our data to suggest improved formulations. Our results thus provide a starting point for a comprehensive parameterisation of topographic stresses in ocean models where fine scale topography is unresolved.}

\begin{document}

\maketitle

%
%
%
\statement
    As the ocean flows over topographic features it experiences a force. Such topography includes undersea mountains that are too small to be accurately described by global ocean and climate models. These forces shape the ocean structure considerably and affect large-scale ocean circulation features, like the Atlantic Meridional Overturning Circulation. Because there is no hope to accurately include these small-scale effects in climate projections we need expressions (a.k.a.\ parametrisations) for the collective force they exert on the ocean. Here, we evaluate existing parametrisations for the force created when an oceanic current flows over an underwater hill by comparing them to the output of small-scale, high-resolution ocean simulations. Compared to previous studies, we cover a much wider range of cases. This allows us to see where existing parametrisations break down and also suggest improvements to these parametrisations, taking us closer to the goal of developing a general and widely applicable mathematical description of this phenomenon.
%
%

%
\section{Introduction}
In shallow seas, fast moving currents are subject to significant energy losses from frictional drag along the ocean bottom. However in the deep ocean, where currents are slower, there is still a significant amount of energy dissipation attributed to the formation of internal waves in regions of rough topography
\citep{ledwell2000evidence,egbert2000significant,egbert2001estimates,naveira2013impact}. These waves can cause substantial mixing, which plays a role in nutrient transport and large scale ocean circulation \citep{sandstrom1984internal,munk1998abyssal,melet2013sensitivity,melet2014sensitivity}.

Over half of the ocean floor is covered by abyssal hills and mountains, defined respectively as having a variation in height of $300-1000$ m and greater than $1000\ \text{m}$ \citep{harris2014geomorphology} over a 750 arc second (or $\approx 10-20\ \mathrm{km}$) radius. Since internal waves generated at these features are of a comparable horizontal scale to the topography itself \citep[e.g.,][]{smith2002conversion}, resolving these waves is not possible with current global ocean models which generally have horizontal resolutions on the order of~10 to 100~km \citep[e.g.,][]{ansong2015indirect,kiss2020access}. As a result, it is important to have parametrisations for the effects of internal waves generated at such hills. Existing parametrisations have mainly focused on computing the energy flux extracted from the flow due to the generation of internal waves \citep[e.g.,][]{simmons2004tidally,jayne2009impact,olbers2013global,de2019toward}. Such a focus is justified, as it has allowed better understanding of the contribution of internal waves to tidal and eddy dissipation, and ocean mixing. However, in recent times, there have been further efforts \citep[e.g.,][]{shakespeare2020drag,klymak2021parameterizing,eden2021closure} which attempt to develop parametrisations for the corresponding topographic stress due to internal waves and other related effects. These effects include blocking \citep{winters2014topographic,nikurashin2014impact} and bottom-trapped tides formed at high latitudes where a stronger Coriolis force prevents internal waves from freely propagating \citep{falahat2015generation,shakespeare2020drag}. Developing these stress parametrisations is vital to obtaining accurate representations of eddies and tides, and large-scale flow features such as the Antarctic Circumpolar Current \citep[e.g.,][]{naveira2013impact,arbic2019connecting}.

To be precise, the energy loss from a spatially uniform flow $\overline{\bm{u}}$ and topographic stress $\boldsymbol{\tau}$ is equal to $\overline{\bm{u}} \cdot \bm{\tau}$, and the topographic stress also acts to modify $\overline{\bm{u}}$. Thus, the topographical stress is a more fundamental quantity to study, in that it can be used to compute energy loss, but this is not necessarily true the other way around. This is particularly the case when $\overline{\bm{u}}$ is time-varying, so that it is not at all valid to use expressions for the time-averaged energy flux to understand the stress $\bm{\tau}$. Doing so omits any component of the stress $\boldsymbol{\tau}$ that is out-of-phase with the flow $\overline{\bm{u}}$, which in some cases is the entire stress \citep[e.g., for bottom-trapped tides;][]{falahat2015generation,shakespeare2020drag}

Due to the complexity of developing a generic parametrisation for stress, existing work \citep[e.g.,][]{bell1975topographically,jayne2001parameterizing,klymak2010high,shakespeare2020drag} has focused on special cases. For instance, previous studies have treated steady and tidal flows in isolation. However, since the steady and tidal components of a flow can interact \citep{shakespeare2020interdependence}, it is important for existing parametrisations to be extended to mixed flows. Existing parametrisations also have several other assumptions that can reduce their accuracy if incorporated into ocean models. For example, it is often assumed that the topography height is small \citep{bell1975topographically, shakespeare2020drag} or that the latitude is close to $0^\circ$ \citep{jayne2001parameterizing,klymak2010high}.

The first goal of this paper is to evaluate the accuracy of a subset of available wave and non-wave stress parametrisations. This is done by comparing theoretical expressions to the stress calculated from a large suite of ocean simulations. The main parametrisations we focus on are those due to \citet{bell1975topographically}, \citet{jayne2001parameterizing}, \cite{klymak2010high} and \citet{shakespeare2020drag}, and we refer to the reader to other related works throughout. Here we focus on the magnitude of the stress the fluid exerts on the topography, rather than discerning exactly where in the water column the equal and opposite reaction stress on the ocean is applied. We also focus on the case of an isolated Gaussian hill in either two or three dimensions, which is a standard, simple case which has been studied in the context of related phenomena \citep{legg2006preliminary,maitland2025oceanic}. A Gaussian hill is a decent model for an isolated undersea mountain (seamount). However, a stronger motivation in this case is that working with isolated hills means that there are only a small number of well-defined parameters to consider (i.e., height and width), allowing us to ensure that the basic physics and scalings of the parametrisations are correct. This is in contrast to other studies \citep[e.g.,][] {jayne2001parameterizing,buijsman2015optimizing,shakespeare2021impact} which tend to only test stress parametrisations with global models or more complicated topography, and often with more limited parameter regimes. In a similar vein, we reduce the complexity of our simulations by taking the stratification to be constant. We also note that whilst three-dimensional simulations are more physically realistic, they are more computationally expensive than two-dimensional simulations. Moreover, two-dimensional topography may be interpreted as approximating a section of a three-dimensional ridge, which can be sites of significant internal wave generation \citep{ledwell2000evidence,garrett2007internal}. Two-dimensional simulations are thus useful despite being less broadly applicable than three-dimensional models. As a result, our general procedure is to run a large number of tests in two dimensions, and then further analyse any interesting phenomena in three dimensions.

The second goal of this paper is to suggest improvements to existing stress parametrisations. This is done using predominantly mathematical or physical arguments that are supported by the simulation data. Notably, one of the common shortcomings in existing parametrisations is the assumption of small topographic height despite tall topographic features being abundant in the ocean \citep{harris2014geomorphology}. Moreover, large isolated seamounts are of interest in ecology, vulcanology and ship navigation \citep{clark2010ecology,watts2019science}. Therefore, it is important for existing parametrisations to be extended to such cases. The ultimate goal of this work is to understand the isolated hill regime in detail and help pave the way for a more complete parametrisation of wave and non-wave stress for implementation in global ocean models.

The structure of the paper is as follows. In Section~\ref{existingsect} we recall some existing drag parametrisations used for tidal, steady and mixed flows. In Section~\ref{modelsect} we discuss the setup for our simulations. In Section~\ref{resultssect} we then discuss the results of the simulations, and use these results to evaluate and suggest improvements to each parametrisation. Finally, in Section~\ref{summsect} we discuss the main takeaways from our analysis, and describe a combined parametrisation that depends on the type of flow and scale of topography under consideration.

\section{Existing parametrisations}\label{existingsect}
We begin by summarising existing parametrisations for wave and non-wave stress. The parametrisations are categorised depending on the flow under consideration --- tidal, steady or mixed. Each parametrisation is expressed in its general form, but we also simplify for the case of an isolated Gaussian hill. For a two-dimensional domain (viewed as cross-section of three-dimensional space) with horizontal component $[-L_x/2,L_x/2]$, we define the stress as
\begin{equation}\label{F2deq}
    F_{\text{2d}}=\frac{1}{\rho_0}\int\limits_{-L_x/2}^{L_x/2} p_h\frac{\mathrm{d}h}{\mathrm{d}x}\mathrm{d}x,
\end{equation}
where $h(x)$ is the bottom topography, $p_h(x)$ the bottom pressure, and $\rho_0$ is the background density of the sea water. In particular, the stress \eqref{F2deq} is the net force per spanwise unit length created from the pressure differential across the hill, divided by $\rho_0$ for convenience. As a result, $F_{\text{2d}}$ has units $\si{N.m^2.kg^{-1}}$ and $\rho_0 F_{\text{2d}}$ has units of force per unit length. The sign of $F_{\text{2d}}$ is also such that a positive value of $F_{\text{2d}}$ corresponds to a force per unit length of magnitude $|\rho_0 F_{\text{2d}}|$ acting on the hill in the positive {$x$-direction}, along with an equal and opposite force acting on the flow in the negative {$x$-direction}. In three dimensions, with a horizontal domain $[-L_x/2,L_x/2]\times[-L_y/2,L_y/2]$, we then define the stress due to a topography $h(x, y)$ as the vector
\begin{equation}\label{F3deq}
    \qquad \bm{F}_{\text{3d}}=\frac{1}{\rho_0}\int\limits_{-L_y/2}^{L_y/2}\int\limits_{-L_x/2}^{L_x/2}  p_h\boldsymbol{\nabla}{h}\:\mathrm{d}x\mathrm{d}y.
\end{equation}
Note that $\bm{F}_{\text{3d}}$ has units $\si{N.m^3.kg^{-1}}$ and $\rho_0 \bm{F}_{\text{3d}}$ has units of force.

\begin{figure*}[h]
    \centering
    \includegraphics[width=0.95\textwidth]{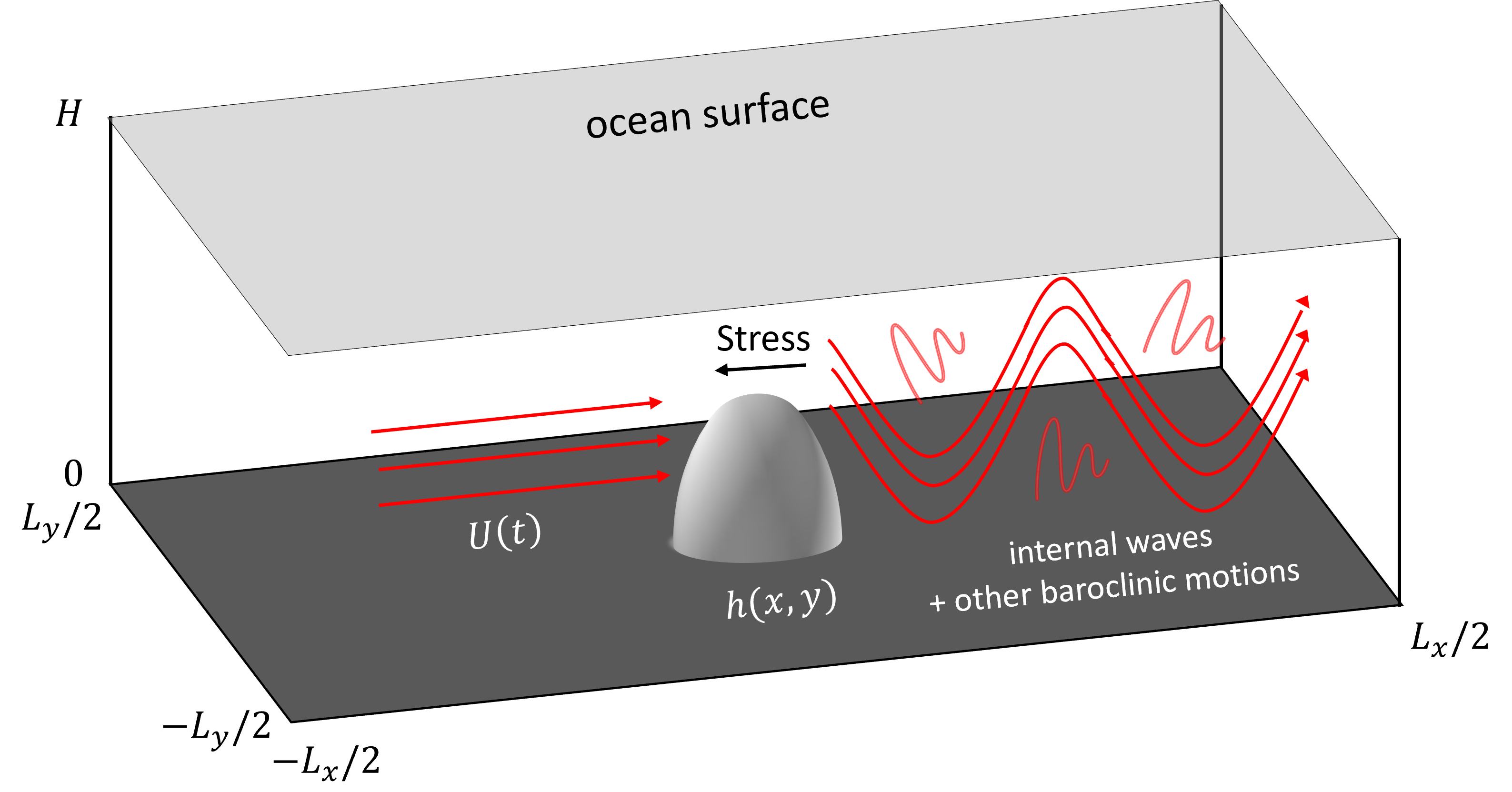}
    \caption{A schematic of the model domain. A spatially mean flow $U(t)$ interacts with the topography $h(x,y)$ to generate baroclinic motions and an associated stress.}\label{fig:domain}
\end{figure*}

In what follows, our domain is as shown in Figure~\ref{fig:domain}, with dimensions $L_x \times L_y \times H$ and a spatially mean zonal flow $U(t)$ over topography $h(x, y)$. For two-dimensional analysis, we restrict ourselves to the $x$--$z$ plane. Moreover, for the ensuing theoretical analysis in this section, we assume $L_x$ and $L_y$ are large. That is, $L_x, L_y\to\infty$.

Throughout this paper, we use the closely related terms ``stress" and ``internal wave drag". To avoid any confusion, we briefly clarify what is meant by each of these terms. By ``stress", or more correctly ``form stress", we mean the total force (scaled by $1/\rho_0$) acting on the flow as per \eqref{F2deq} or \eqref{F3deq}.  ``Internal wave drag" refers to the component of form stress due to the formation of internal waves. Many existing parametrisations focus only on internal wave drag, which is the dominant component of stress in many idealised scenarios, but not if the flow is blocked or sufficiently non-linear.

\subsection{Case 1: Tidal Flow}\label{secttidalparam}
We consider a periodic tidal flow $U(t)=U_\text{tidal}\cos(\omega t)$ in the $x$-direction where $U_{\text{tidal}}$ is a constant and $\omega$ is the tidal frequency. Tidal flows are a major source of internal waves and the stress associated with their generation impacts the strength of both barotropic and internal tides \citep{ansong2015indirect,buijsman2015optimizing}. As a result, this flow regime is the main focus in the oceanic literature.

In the work of \citet{bell1975topographically}, linear theory was used to develop an expression for the energy flux generated by internal tidal waves. Bell found his theory to compare favorably with observations, and his parametrisations have since been refined by others including \citet{smith2002conversion}, \citet{khatiwala2003generation} and \citet{nycander2005generation}. Although Bell's theory was derived for small-scale topography, recent work \citep{geoffroy2024tidal,pollmann2023resolving} suggests that Bell's linear theory can still be reliable for moderately large topography, particularly for low mode waves. However, because Bell's parametrisation and its derivatives provide expressions for the time-averaged energy flux, they cannot be directly converted to time-dependent stress parametrisations.

For our analysis, we focus on two easily applied stress parametrisations closely related to the work of \citet{bell1975topographically}. First, we will discuss an older and ubiquitously applied parametrisation due to \citet[hereafter, JSL2001]{jayne2001parameterizing}, which is essentially a scaling for stress based on Bell's formula for the internal tide energy flux. Then, we will analyse a more mathematically rigorous parametrisation due to \citet[hereafter SAH2020]{shakespeare2020drag}. A comparison of other similar parametrisations, with a focus on energy conversion, is given by \citet{green2013comparison}.

\subsubsection{Jayne and St Laurent parametrisation}
Jayne and St Laurent give a simple estimate for internal wave drag, by assuming that the stress scales in an identical way to the Bell's energy flux formula. Despite the simple form of the JSL2001 parametrisation, it has been shown to significantly improve the accuracy of tidal models when compared to observations, and is widely used today \citep{arbic2018primer,ansong2015indirect}. For each grid cell, they set the stress (in each horizontal dimension) to be proportional to $Nh^2u$. Here, $N$ is the buoyancy frequency, and $h$ and $u$ are taken to be suitable values of the topography height and tidal velocity in the grid cell. For small $h\ll H$, we can approximate $u$ as the barotropic flow $U(t)$ and $N$ as a constant in the vicinity of the topography. Thus, the total stress over a small hill with root mean square height $h_{rms}$ is estimated by
\begin{equation}\label{force-jsl}
    F_{\text{JSL}}=\frac{1}{2}\kappa N_Bh^2_{rms}U.
\end{equation}
Here, $\kappa$ is the tunable scaling constant in JSL2001, $N_B$ denotes the buoyancy frequency at the seabed, and the subscript ``JSL" indicates that this is the drag force due to the JSL2001 theory. In line with our units and scaling for stress in \eqref{F2deq} and \eqref{F3deq}, here $\kappa$ has units of length in three dimensions and is unitless in two dimensions. On the other hand, in JSL2001 the stress is implicitly divided by the horizontal length (in two-dimensions) or area (in three-dimensions) so that $\kappa$ has units of inverse length and may therefore be interpreted as the characteristic wavenumber of the topography. Note that the factor of $1/2$ in \eqref{force-jsl} could be absorbed into the definition of $\kappa$, but we have included it to more closely match the expression given by JSL2001.

Although simple and widely used, there are some notable limitations of the JSL2001 model. Firstly, the model was only designed for small-scale topography and large ocean depths. In addition:
\begin{enumerate}
    \item Representing a complicated region of topography by a single wavenumber $\kappa$ is very simplistic and a more sophisticated formulation would be preferred.
    \item Using the bottom value of stratification $N_B$ has since been shown to be unreliable, and it is more accurate to use an averaged form of the stratification \citep{zarroug2010energetics,luschow2024sensitivity}.
    \item The parametrisation \eqref{force-jsl} is latitude-independent, an omission which Jayne and St Laurent state could be significant when the tidal frequency $\omega$ is close to the Coriolis frequency $f$.
\end{enumerate}

\subsubsection{Shakespeare, Arbic and Hogg parametrisation}\label{SAHtheorysect}
In SAH2020, the authors take a more analytic approach to parametrising tidal flow stress, giving an expression which refines that of JSL2001 (Equation \eqref{force-jsl}). The main novelty of their work is their treatment of ``spring" forces, such as bottom-trapped tides, which are out-of-phase with the tidal flow and so do not do any work in the time mean. This is important, because JSL2001 and other prior parametrisations compute stress via a comparison to the time-averaged energy flux, and thus do not account for such spring forces. Since the parametrisation of SAH2020 refines that of JSL2001, including by adding frequency dependence (point 3 above) and a more sophisticated representation of the topographic spectrum (1 above), we mainly compare our numerical simulations with the SAH2020 theory, whilst commenting on the limitations of the JSL2001 theory.

Under standard assumptions, including small topographic height and large ocean depth, SAH2020 give the stress for subcritical latitudes ($|f|<\omega$) and isotropic topography as
\begin{equation}\label{SAH-forcefull}
    F_{\text{3d}}=\frac{\sqrt{(N^2-\alpha\omega^2)(\omega^2-f^2)}}{4\pi |\omega|}U\int_0^\infty|\hat{h}(K)|^2K^2\mathrm{d}K.
\end{equation}
Here, $\alpha\in\{0,1\}$ is such that
\begin{equation}\label{alphadef}
    \alpha=
    \begin{cases}
        0,&\text{under the hydrostatic assumption},\\
        1,&\text{otherwise.}
    \end{cases}
\end{equation}
Then, $\hat{h}$ denotes the Fourier transform of $h$ in $t$, $x$ and $y$ to transformed variables $\omega$, $k$ and $\ell$ respectively with $K \equiv \sqrt{k^2+\ell^2}$. Note that we use the Fourier transform convention
\begin{equation}
    \hat{g}(k) \equiv \int_{-\infty}^{\infty} g(x) \, e^{-ikx}\mathrm{d}x.
    \label{eq:Fourierconvention}
\end{equation}
Since we are assuming the topography is isotropic (and independent of time), $\hat{h}$ is a function of the total wavenumber $K$ only, as indicated in \eqref{SAH-forcefull}. We also remark that \eqref{SAH-forcefull} is only the $x$-component of the stress, as this is the primary direction of our flow and focus of our analysis.

As a direct comparison to the JSL2001 parametrisation, a clear frequency dependency has been added and the constant $\kappa$  in \eqref{force-jsl} has been replaced with the analytic expression
\begin{equation*}
    \frac{1}{2 \pi h_{\text{rms}}^2}\int_0^\infty|\hat{h}(K)|^2K^2\mathrm{d}K.
\end{equation*}
In particular, in the limit where $N\gg \omega\gg f$, $F_{\text{3d}}$ reduces exactly  to the JSL2001 expression \eqref{force-jsl}, albeit with an analytic expression for what $\kappa$ should be set to.

In SAH2020, a more general expression than \eqref{SAH-forcefull} is stated which accounts for wave reflections off the ocean surface. However, such reflections are difficult to parametrise \citep[e.g.,][]{shakespeare2021dissipating}, and for a narrow, isolated hill in a deep ocean the number of waves reflecting back onto the topography is small. Therefore, we omit any reflection terms from \eqref{SAH-forcefull} for simplicity. Nevertheless, we note that these reflections become important even for an isolated hill when the half-vertical wavelength $\lambda_v$ compared to the ocean depth $H$, known as the mode-number
\begin{equation}\label{vwaveeq}
    n=\frac{H}{\lambda_v/2}=\frac{K H\sqrt{N^2-\alpha\omega^2}}{\pi\sqrt{\omega^2-f^2}},
\end{equation}
approaches 1 (from above). As a result, we expect \eqref{SAH-forcefull} may have deviations from the actual stress for small $K$, $H$ or $N$.

In the case of an isolated Gaussian hill with width $W$, whereby
\begin{equation}
    h(x,y)=h_0 e^{-(x^2+y^2)/(2W^2)}
\end{equation}
we obtain from \eqref{SAH-forcefull}
\begin{equation}\label{SAH3dgauss}
    F_{\text{3d}}=\frac{\pi\sqrt{\pi}}{4}W \frac{\sqrt{(N^2-\alpha\omega^2)(\omega^2-f^2)}}{\omega}h_0^2 U.
\end{equation}
Importantly, \eqref{SAH3dgauss} suggests that the stress depends linearly on the hill width $W$, at least for Gaussian hills.

The expression for supercritical (bottom-trapped) latitudes $|f|\geq\omega$ is similar, but with the stress out of phase with velocity. In particular, for a Gaussian hill
\ifdraft
\begin{equation}\label{SAH3d}
    F_{\text{SAH3d}}=
    \begin{cases}
        \frac{\pi\sqrt{\pi}}{4}W\omega^{-1}\sqrt{(N^2-\alpha\omega^2)(\omega^2-f^2)}\,h_0^2\, U_{\text{tidal}}\cos(\omega t),& |f|<\omega,\\
        \frac{\pi\sqrt{\pi}}{4}W \omega^{-1}\sqrt{(N^2-\alpha\omega^2)(f^2-\omega^2)}\,h_0^2\, U_{\text{tidal}}\sin(\omega t),& |f|\geq\omega,
    \end{cases}
\end{equation}
\else
\begin{align}\label{SAH3d}
    &F_{\text{SAH3d}}= \frac{\pi\sqrt{\pi}}{4}W\omega^{-1} \,h_0^2 U_{\text{tidal}} \times\nonumber\\
    &\ \times
    \begin{cases}
       \sqrt{(N^2-\alpha\omega^2)(\omega^2-f^2)}\, \cos(\omega t),\ |f|<\omega,\\
        \sqrt{(N^2-\alpha\omega^2)(f^2-\omega^2)}\, \sin(\omega t),\ |f|\geq\omega,
    \end{cases}
\end{align}
\fi
where the subscript SAH3d indicates that this is the three-dimensional stress due to the SAH2020 theory. In a subsequent paper \citep{shakespeare2021impact}, the authors also give a parametrisation for two dimensions. Namely, for a Gaussian hill $h(x) = h_0\,e^{-x^2/(2W^2)}$,
\ifdraft
\begin{equation}\label{SAH2d}
    F_{\text{SAH2d}}=
    \begin{cases}
        \omega^{-1}\sqrt{(N^2-\alpha\omega^2)(\omega^2-f^2)}\,h_0^2 \,U_{\text{tidal}}\cos(\omega t),& |f|<\omega,\\
        \sqrt{(N^2-\alpha\omega^2)(f^2-\omega^2)}\,h_0^2 \,U_{\text{tidal}}\sin(\omega t),& |f|\geq\omega.
    \end{cases}
\end{equation}
\else
\begin{align}\label{SAH2d}
    &F_{\text{SAH2d}}= \omega^{-1}\,h_0^2\, U_{\text{tidal}} \times \nonumber \\
    &\ \times \begin{cases}
        \sqrt{(N^2-\alpha\omega^2)(\omega^2-f^2)} \cos(\omega t),& |f|<\omega,\\
        \sqrt{(N^2-\alpha\omega^2)(f^2-\omega^2)}\sin(\omega t),& |f|\geq\omega.
    \end{cases}
\end{align}
\fi
Note that~\eqref{SAH2d} is identical to~\eqref{SAH3d} except without the factor of $\pi\sqrt{\pi} W/4$. To obtain \eqref{SAH3d} and \eqref{SAH2d} one also needs to assume that $N$ is constant, which we will do throughout our analysis. As discussed before, in the more physical setting where $N$ is varying function of depth, it is better to use an averaged value of $N$, at least for sufficiently large hill widths  \citep{luschow2024sensitivity}.

\subsection{Case 2: Steady Flow}\label{steadytheorysect}
We now consider a constant steady flow $U$ in the $x$-direction. Under such a flow, the internal waves formed are called lee waves. These lee waves have zero Eulerian frequency (i.e., are steady at a fixed point), and are formed with intrinsic frequency depending on the topography and the magnitude of $U$. This leads to some added complexities compared to the tidal case, where the internal waves have the same intrinsic frequency as the tide. In particular, for steady flows there is a greater dependency on the Froude number
\begin{equation}
    Fr \equiv \frac{Nh}{U}.
\end{equation}
For large Froude numbers, the system becomes highly nonlinear and harder to parametrise. Namely, when $Fr\gtrsim 1$, this non-linearity manifests as blocking or splitting, which can significantly decrease the formation of lee waves \citep{nikurashin2014impact}.

As in the tidal case, we also focus on two parametrisations for steady flows, treating the cases of small and large Froude numbers separately. The first, which is based on the linear theory of \citet{bell1975topographically}, is an idealised theory assuming lee waves form without the effects of a high Froude number. The second, is a semi-empirical parametrisation due to \citet[hereafter, KLP2010]{klymak2010high}. The KLP2010 parametrisation focuses instead on a regime where most of the flow is blocked, and unable to surmount the topography to form lee waves. These two parametrisations therefore look at different aspects of the wave and non-wave stress, and so both are useful in obtaining a wholistic parametrisation. Of course, there is also the transitional case where Froude number is of a moderate size, say $Fr\approx 1-10$. To limit the scope of our analysis we do not explore this case in depth here, but refer the reader to recent attempts at developing a hybrid parametrisation \citep{perfect2020energetics,klymak2021parameterizing,baker2022impact}.

\subsubsection{Bell parametrisation for steady flows}\label{bellsteadysect}
In the work of \citet{bell1975topographically}, an expression for the time-mean stress due to lee waves is provided. In three dimensions, his expression for the $x$-component of stress is
\ifdraft
\begin{align}\label{bell-full3d}
    F_{\text{Bell3d}}=\frac{1}{4\pi^2}\iint\frac{k}{\sqrt{k^2+\ell^2}}|\hat{h}(k,\ell)|^2\sqrt{[N^2-\alpha(kU)^2][(kU)^2-f^2]}\mathrm{d}k\mathrm{d}\ell
\end{align}
\else
\begin{align}\label{bell-full3d}
    &F_{\text{Bell3d}}=\nonumber\\
    &\ =\frac{1}{4\pi^2}\iint\frac{k\, |\hat{h}(k,\ell)|^2}{\sqrt{k^2+\ell^2}}\sqrt{[N^2-\alpha(kU)^2][(kU)^2-f^2]}\mathrm{d}k\mathrm{d}\ell
\end{align}
\fi
where $\alpha$ is as in \eqref{alphadef} and the integrals are taken over all real values of $k$ and $\ell$ under the restriction ${|f|\leq |kU|<\infty}$ for hydrostatic ($\alpha=0$) and $|f|\leq |kU|\leq N$ for non-hydrostatic ($\alpha=1$) conditions. As we are taking the flow speed to be constant, the lee wave stress is non-zero in the time-mean and there is not an equivalent bottom-trapped spring force as in the SAH2020 tidal parametrisation theory. However, it is worth noting that \eqref{bell-full3d} can readily be obtained by repeating the arguments in SAH2020 for a steady flow. The two-dimensional equivalent of \eqref{bell-full3d} is then
\ifdraft
\begin{equation}\label{bell-full2d}
    F_{\text{Bell2d}}=\frac{1}{2\pi}\int\limits_{\substack{|f|\leq|kU|\\
    \alpha|kU|\leq N}}|\hat{h}(k)|^2\sqrt{[N^2-\alpha(kU)^2][(kU)^2-f^2]}\,\mathrm{d}k.
\end{equation}
\else
\begin{align}\label{bell-full2d}
    &F_{\text{Bell2d}}=\nonumber\\
    &\ =\frac{1}{2\pi}\int\limits_{\substack{|f|\leq|kU|\\
    \alpha|kU|\leq N}}|\hat{h}(k)|^2\sqrt{[N^2-\alpha(kU)^2][(kU)^2-f^2]}\,\mathrm{d}k.
\end{align}
\fi

We now consider the case of an isolated Gaussian hill of height $h_0$. In general, \eqref{bell-full3d} and \eqref{bell-full2d} do not simplify much further unless one wishes to express the parametrisations in terms of incomplete Bessel functions. However, in the special case at the equator ($f=0$) with a hydrostatic assumption ($\alpha=0$), we have
\begin{align}
    F_{\text{Bell3d}}(\alpha=f=0)&=\frac{\pi\sqrt{\pi}}{4}W Nh_0^2 U\label{bell3dlim},\\
    F_{\text{Bell2d}}(\alpha=f=0)&= Nh_0^2 U\label{bell2dlim},
\end{align}
which is the same as what one obtains in the tidal case (cf.\ \eqref{SAH3d} and \eqref{SAH2d}).

Bell's parametrisations are derived in the case $Fr\ll 1$. For higher Froude numbers the lee wave flux saturates as non-linear blocking and splitting effects take hold \citep{nikurashin2014impact}. In more recent energy flux parametrisations \citep[e.g.,][]{baker2022impact}, a simple scaling is used to account for this phenomena. Namely, let $E$ denote the internal wave energy flux and $E_{\text{Bell}}$ denote the energy flux predicted by the \citet{bell1975topographically} theory. Then for some critical Froude number $Fr_c$ one sets
\begin{equation}\label{frcriteq}
    E=\left(\frac{Fr_c}{Fr}\right)^2E_{\text{Bell}}.
\end{equation}
whenever $Fr\geq Fr_c$. The standard values of $Fr_c$ used are 0.7 in two dimensions, and 0.4 in three dimensions. These values were derived empirically from experimental observations \citep{aguilar2006internal} and simulation data \citep{nikurashin2014impact}.

\subsubsection{Klymak, Legg and Pinkel parametrisation}\label{Klysect}
In KLP2010, a form stress parametrisation is obtained for when the Froude number $Fr$ is very large, whereby blocking effects are the dominant form of stress. Their parametrisation is derived using numerical simulations for two-dimensional Gaussian hills under the hydrostatic assumption and $f=0$. The full expression for the form stress in KLP2010 is
\begin{equation}\label{KLPforce}
    F_{\text{KLP}}=Nh_0^2 U_m\left[1+\pi\frac{U_m}{Nh_0}-2\pi^2\left(\frac{U_m}{Nh_0}\right)^2\right]
\end{equation}
where $h_0$ is the hill height and
\begin{equation}\label{Umsteadydef}
    U_m=\frac{H}{H-h_0}U.
\end{equation}
In KLP2010, a factor of $\pi/2$ actually appears out the front of \eqref{KLPforce}. However, this is the result of a small algebraic error so that the factor is actually $\pi/3\approx 1$.

For $Fr\gg 1$, the second and third terms in \eqref{KLPforce} are small, so that this parametrisation is nearly equivalent to the linear lee wave stress given in \eqref{bell2dlim}. The main difference is the scaling of $U$ by $H/(H-h)$, which represents the increase in velocity as the flow moves over the narrower channel above the hill. This suggests that when $f=0$, an approximate $Nh_0^2U$ scaling could be accurate for both low and high Froude numbers.

To limit the scope of this paper we focus on the cases $Fr\ll 1$ (\citet{bell1975topographically} theory) and $Fr\gg 1$ (KLP2010 theory) discussed above. However, there are also intermediate regimes $Fr\approx 1$ that one can consider. In this regard, ``hybrid" parametrisations have been developed in both the atmospheric \citep{garner2005topographic,lott1997new} and oceanic \citep{klymak2021parameterizing,perfect2020energetics} literature. This style of parametrisation has been successfully compared with both global models \citep{trossman2013impact,trossman2016impact} and observations \citep{trossman2015internal}. Such parametrisations generally represent the stress as the sum of a propragating (wave) component near the crest of the topography, and a non-propagating (blocked) component at the base of the topography.

\subsection{Case 3: Mixed Flow}\label{mixedtheorysect}
Finally, we consider a mixed flow in the $x$-direction of the form
\begin{equation}\label{Uequation}
    U(t)=U_{\text{mean}}+U_{\text{tidal}}\cos(\omega t),
\end{equation}
where $U_{\text{mean}}$ and $U_{\text{tidal}}$ are constants. Compared to the tidal and steady cases, very little work has been done on mixed flows. We thus hope that our subsequent simulations can shine a light on previously unconsidered regimes.

A priori, we expect that the stress for a mixed flow should have a mean ($F_m$) and periodic ($F_p$) component i.e.
\begin{equation}\label{F3dsplit}
    F_{3d}=F_m+F_p\cos(\omega t+\phi)
\end{equation}
where $\phi$ denotes a phase shift. A naive parametrisation would be to assume that the mean (steady) and tidal components of the flow do not interact \citep[although we know this is not the case; e.g., ][]{shakespeare2020interdependence}. Nonetheless, if we assume that $F_m$ is independent of $U_{\text{tidal}}$ and $F_p$ is independent of $U_{\text{mean}}$, then $F_m$ and $F_p$ can be directly computed from the parametrisations in the steady and tidal cases respectively. There does not appear to be any well-developed parametrisation for $F_p$ in the literature \citep[][only consider the energy flux]{bell1975topographically,shakespeare2020interdependence}, so we primarily compare our simulation data to this naive parametrisation for $F_p$. Certainly though, one should expect some impact of the mean flow
on the periodic stress given that the mean flow will act to
Doppler shift the tidal frequency. For the mean component $F_m$ however, there is some existing theory available, which we now discuss.

\subsubsection*{Bell parametrisation for mixed flows}\label{bellmixedsect}
Although we previously discussed the \citet{bell1975topographically} theory for steady flows, Bell also gives a more general expression for the mean stress in mixed flows. \citet{shakespeare2020interdependence} extended this expression to include multiple tidal frequencies, but we will only consider the simpler Bell formulation here. In three dimensions, Bell gives the following formula for the $x$-component of the time-mean stress (cf. \eqref{bell-full3d})
\ifdraft
\begin{align}\label{bellmixed}
    F_{m,3\text{d}}=\frac{1}{4\pi^2}\sum_{n=-\infty}^\infty\iint\frac{k}{\sqrt{k^2+\ell^2}}|\hat{h}(k,\ell)|^2\sqrt{(N^2-\alpha\omega_n^2)(\omega_n^2-f^2)} \, J_n^2\left(\frac{kU_{\text{tidal}}}{\omega}\right)\mathrm{d}k\mathrm{d}\ell,
\end{align}
\else
\begin{align}\label{bellmixed}
    &F_{m,3\text{d}}=\nonumber\\
    &\ =\frac{1}{4\pi^2}\sum_{n=-\infty}^\infty\iint\frac{k\, |\hat{h}(k,\ell)|^2}{\sqrt{k^2+\ell^2}}\sqrt{(N^2-\alpha\omega_n^2)(\omega_n^2-f^2)} \, \times \nonumber\\
    &\qquad\qquad\qquad\qquad\times J_n^2\left(\frac{kU_{\text{tidal}}}{\omega}\right)\mathrm{d}k\mathrm{d}\ell,
\end{align}
\fi
where $\omega_n=n\omega+kU_{\text{mean}}$, $J_n$ is the $n$th-order Bessel function of the first kind, and the integrals are taken over all real values of $k$ and $\ell$ under the restriction $|f|\leq|\omega_n|<\infty$ if $\alpha=0$ and $|f|\leq|\omega_n|\leq N$ if $\alpha=1$. The two-dimensional equivalent, which we denote $F_{m,2d}$ is then given by
\ifdraft
\begin{equation}\label{bellmixed2d}
    F_{m,2\text{d}}=\frac{1}{2\pi}\sum_{n=-\infty}^\infty\int\limits_{\substack{|f|\leq|\omega_n|\\
    \alpha|\omega_n|\leq N}}|\hat{h}(k)|^2\sqrt{(N^2-\alpha\omega_n^2)(\omega_n^2-f^2)} \, J_n^2\left(\frac{kU_{\text{tidal}}}{\omega}\right)\mathrm{d}k.
\end{equation}
\else
\begin{align}\label{bellmixed2d}
    & F_{m,2\text{d}}=\nonumber\\
    & \ = \frac{1}{2\pi}\sum_{n=-\infty}^\infty\int\limits_{\substack{|f|\leq|\omega_n|\\
    \alpha|\omega_n|\leq N}}|\hat{h}(k)|^2\sqrt{(N^2-\alpha\omega_n^2)(\omega_n^2-f^2)} \, \times \nonumber\\
    &\qquad\qquad \qquad \qquad \times J_n^2\left(\frac{kU_{\text{tidal}}}{\omega}\right)\mathrm{d}k.
\end{align}
\fi

There are some notable features of \eqref{bellmixed}. Firstly, if $U_{\text{tidal}}=0$ then \eqref{bellmixed} reduces to the same expression in the steady flow case \eqref{bell-full2d}. Moreover, if $U_{\text{mean}}=0$, we have $F_{m,3\text{d}}=F_{m,2\text{d}}=0$. We also note that, provided $x$ is not too large, $J_0(x)$ decreases from $1$ whilst $J_n(x)$ increases from $0$ for $n\geq 1$. This can be interpreted as the lee wave energy ($n=0$) being transferred to higher modes ($n\geq 1$) in the presence of a tidal flow \citep{shakespeare2020interdependence}. However, in most cases $J_n(x)$ ($n\geq 1$) is sufficiently small so that the time-mean stress predicted by \eqref{bellmixed} and \eqref{bellmixed2d} is not too different from the case of a purely steady flow.

Like the Bell parametrisation for a steady flow, \eqref{bellmixed} and \eqref{bellmixed2d} are derived in the linear regime, where the Froude number is assumed to be small. Thus, for large Froude numbers, where this theory is invalid, we will primarily compare our simulation data to the naive parametrisation which assumes that the tidal and mean components of the flow do not interact.

\section{Numerical Model}\label{modelsect}
We use the Julia library \textit{Oceananigans} \citep{OceananigansJOSS, Oceananigans-overview} to model flow over an isolated Gaussian hill in two or three dimensions. Importantly, the GPU-capability of Oceananigans allows us to run several hundred simulations whilst using only a modest amount of computational resources (see also \cite{silvestri2024gpu}). In what follows, we focus on the setup in three dimensions, as the two-dimensional simulations are essentially the same but with only one grid cell in the $y$-direction and improvements in the resolution. For more precise details of the setup see \citet{johnston_2025_17274188}, which contains the base code used for each simulation.

By default, our domain was set to be $L_x=L_y=500\ \text{km}$ long and wide, $H=1.5\ \text{km}$ deep and centred on a symmetric Gaussian hill $h(x,y) = h_0\,e^{-(x^2+y^2)/(2W^2)}$ with height $h_0$ and width $W=5\ \text{km}$. Our domain was periodic in the horizontal $x$-$y$ directions, and bounded in the vertical. To capture the features near the hill, the domain grid was stretched to give a horizontal spacing of $\Delta x=\Delta y \approx 700\ \text{m}$ around the hill, smoothly varied to $\Delta x = \Delta y \approx 2100\ \text{m}$ at the edge of the domain. Similarly, the vertical grid spacing was set to be $\Delta z\approx 7.5\ \text{m}$ in the bottom half of the domain, smoothly varied to $\Delta z\approx 22.5\ \text{m}$ near the surface. The uneven bathymetry is implemented in Oceananigans as an immersed boundary.

To keep the hill isolated, we artificially damped wave features at the horizontal boundaries by introducing a ``sponge layer" in the horizontal regions $[-L_x/2,-L_x/4]$, $[L_x/4,L_x/2]$, $[-L_y/2,-L_y/4]$ and $[L_y/4,L_y/2]$. The sponge layer, in $x$, is:
\begin{equation}
    \mathrm{sponge}(x) = \frac1{2} \mathrm{Erf}\left(\frac{x-L_x/4}{L_x/16}\right) + \frac1{2} \mathrm{Erf}\left(\frac{-x-L_x/4}{L_x/16}\right) + 1,
\end{equation}
and similarly in the $y$-direction for three-dimensional simulations. The sponge region includes a combination of enhanced explicit horizontal diffusivity and horizontal viscosity (peaking at $2000\ \text{m}^2\text{s}^{-1}$ in the outer regions of the domain) and a relaxation of both buoyancy and horizontal velocity to the background state at a rate of $1 / (1.38\;\mathrm{days})$. To verify that the position of the sponge layer did not affect our results, we also ran initial tests (not shown) with a wider domain, and found negligible change in the output. Implicit dissipation of grid-scale kinetic energy is provided by the advection scheme used, which is an upwind-biased, nominally 9th-order Weighted, Essentially Non-Oscillatory advection scheme, commonly known as WENO \citep{pressel2017numerics, shu2020essentially, silvestri2024new}.

For each simulation, we specified constant values for background stratification~($N$), hill height~($h_0$), hill width~($W$), ocean depth~($H$), Coriolis frequency~($f$), and tidal frequency~($\omega$) if required. Depending on whether we were concerned with a tidal, mean or mixed case, we also specified the flow speeds $U_{\text{mean}}$ and $U_{\text{tidal}}$ (see~\eqref{Uequation}).
The velocity is initialised as
\begin{equation}
\boldsymbol{u}(t=0) =(U_{\text{mean}}+U_{\text{tidal}}) \hat{\boldsymbol{x}} ,
\end{equation}
and buoyancy is initialised with a linear constant stratification:
\begin{equation}
    b(t=0) = N^2 z .
\end{equation}
The horizontal momentum equations are forced analogously to the simulations by \citet{klymak2018nonpropagating} and \citet{shakespeare2021impact} to maintain a background zonal velocity ${U(t)=U_{\text{mean}}+U_{\text{tidal}}\cos(\omega t)}$. In particular, the $x$-momentum equation is forced with:
\begin{equation}
    \frac{f^2-\omega^2}{\omega}U_{\text{tidal}} \sin(\omega t) ,
\end{equation}
and the $y$-momentum equation with $f U_{\text{mean}}$.

Our general procedure was to run a test using reference values for $N,\ h_0,\ W,\ H,\ f,\ \omega,\ U_{\text{mean}}$ and $U_{\text{tidal}}$, before varying each parameter individually to determine their impact on the stress. For the purely tidal or steady flow tests, these default parameters are listed in Table~\ref{paramtable}. Note that the bottom-trapped tests are run at $90^\circ S$ which, although unphysical, is fine in a simulation and allows us to study a regime whereby bottom-trapped effects truly dominate. For the mixed flow case, the parameters will either be the same as the tidal case with a mean velocity of $U_{\text{mean}}=0.1\ \si{m.s^{-1}}$ added, or the same as the mean case with a tidal velocity of $0.1\cos(\omega t)\ \si{m.s^{-1}}$ added. This will be made clear in the text. Also note that the default values of $h_0$ and $H$ vary slightly between the two- and three-dimensional tests. This is due to it being more difficult to achieve high resolution for small hills in the three-dimensional case.

In each test, the main quantity we computed was the $x$-component of the stress, given as in \eqref{F3deq} by:
\begin{equation}\label{F3deq2}
    \qquad F_{\text{3d}}=\frac{1}{\rho_0}\int\limits_{-L_y/2}^{L_y/2}\int\limits_{-L_x/2}^{L_x/2} p_h\frac{\partial{h}}{\partial x}\mathrm{d}x\mathrm{d}y,
\end{equation}
where $p_h(x,y)$ is the bottom pressure, i.e., the pressure values that corresponds to the grid cell directly above the hill at horizontal position $(x, y)$.

The default values in columns~1 and~2 of Table \ref{paramtable} were chosen to be within the usual linear regimes. This allowed us to clearly see any non-linear effects that arose when the parameters are varied. For the tidal tests in column~1, a standard measure of non-linearity for propagating waves with $|f|<\omega$ is the steepness parameter, defined by (see e.g.~\cite{balmforth2002tidal})
\begin{equation}\label{steepeq}
    s = k h_0 \frac{\sqrt{N^2-\alpha\omega^2}}{\sqrt{\omega^2-f^2}}.
\end{equation}
For the default values in Table \ref{paramtable} with $|f|=5 \times 10^{-5}$ and $\omega=1.4 \times 10^{-4}$, setting $k$ to be the dominant wavenumber ($k=1/W$) gives $s\approx 0.6$ which is within the ``subcritical" regime $s\lesssim 1$. Then, for the default steady tests in column~2 of Table~\ref{paramtable}, the Froude number is given by $Fr=0.2$ for the two-dimensional tests, and $Fr=0.3$ for the three-dimensional tests, both of which are well within the linear regime.

For efficiency, all simulations were run in hydrostatic mode, using the \textit{HydrostaticFreeSurfaceModel} in Oceananigans. The unimportance of non-hydrostatic processes for the stress is reflected by the parametrisations in Section~\ref{existingsect}, whereby the hydrostatic ($\alpha=0$) and non-hydrostatic ($\alpha=1$) parametrisations are essentially the same provided the buoyancy frequency $N$ is large compared to the wave frequency. More precisely, non-hydrostatic effects are deemed unimportant for $N\gg\omega$ for tidal flows, and $N\gg kU$ for mean flows, which holds for the parameters considered here. Each three-dimensional tidal test was run for 200 hours, whereas each three-dimensional steady test was run for 100 hours. We found that 100 hours was generally long enough to reach a near-equilibrium state, with the slightly longer test time given for tidal tests as to have more full tidal cycles to perform a statistical analysis on. Since two-dimensional tests were much less computationally expensive, we decided to run them for even longer (750 hours) to improve accuracy and detect any notable effects that only appeared at later times. We timestep all simulations using a 2nd-order quasi Adams-Bashforth time stepping scheme. The free surface was substepped with the split-explicit free surface solver described by \citet{SHCHEPETKIN2005347}. All simulations used time-step of~7.5 minutes, and an output interval of 15 minutes.

\section{Results and Evaluation}\label{resultssect}
We now compare the parametrisations in Section~\ref{existingsect} with the numerical data from the model described in Section~\ref{modelsect}. We consider results separately depending on the type of flow (tidal, steady or mixed) and the number of dimensions (two or three).

For each of the purely tidal flow simulations, we assume that (the $x$-component of) the stress is periodic and of the form
\begin{equation}\label{sinesteq}
    \text{Stress}(t)= A\cos(\Omega t + \phi),
\end{equation}
for some determinable values of the amplitude $A$, frequency $\Omega$ and phase $\phi$. To compute these values, we first extract the last 200 values (corresponding to 40 hours) of the stress output by the model, computed as per \eqref{F2deq} or \eqref{F3deq2}, depending on the dimension under consideration. We then fit these 200 values to a cosine curve. To calculate uncertainty, we recompute these values for the last 100 values and the last 101--200 values, recording the difference from the initially obtained values.

For each of the purely steady flow simulations, we use the average value of the stress over the last 20 hours of simulation. The uncertainty is then computed using the minimum and maximum values of the stress during these last 20 hours.

Finally, for the mixed flow simulations, we assume the stress is of the form (cf.~\ \eqref{F3dsplit})
\begin{equation}\label{sinestmix}
    \text{Stress}(t)= F_m+F_p\cos(\Omega t + \phi),
\end{equation}
with both a mean component $F_m$ and a periodic component $F_p\cos(\Omega t+\phi)$. To compute $F_m$ and $F_p$ we fit the last 200 output values of the stress to a shifted cosine curve of the form \eqref{sinestmix}. The uncertainty in $F_m$ and $F_p$ is obtained by recomputing for the last 100 values and 101--200 values and recording any difference.

\def\arraystretch{1.5}
\begin{table*}[h]
\centering
\caption{The default parameter values used for the tests with a purely tidal or steady flow in either two dimensions~(2d) or three dimensions~(3d). Here $N$ is the buoyancy frequency, $h_0$ is the hill height, $W$ is the hill width, $H$ is the ocean depth, $f$ is the Coriolis parameter, $\omega$ is the tidal frequency, $U_{\text{mean}}$ is the mean flow speed and $U_{\text{tidal}}$ is the amplitude of the periodic component of the flow. For the tidal flow tests, the choice of $\omega$ and $f$ (and thus latitude) will be made clear in the text. The parameters for the steady flow tests also vary depending on whether we are interested in a low or high Froude number $Fr=Nh_0/U_{\text{mean}}$ regime.}
\begin{tabular}{|c|c|c|c|}
\hline
& Tidal & Steady $(Fr\lesssim 1)$ & Steady ($Fr\gg 1$)\\
\hline
$N$ $(\mathrm{s}^{-1})$ & $4\times 10^{-3}$ & $2\times 10^{-3}$ & $5\times 10^{-3}$\\
\hline
$h_0$ $(\mathrm{m})$ & $100$ & $20$ (2d) or $30$ (3d) & $500$\\
\hline
$W$ $(\mathrm{m})$& $5000$ & $5000$ & $5000$\\
\hline
$H$ $(\mathrm{m})$& $1500$ & $1500$ (2d) or $1250$ (3d) & $1500$\\
\hline
$f$ $(\mathrm{s}^{-1})$& $-5\times 10^{-5}$ or $-1.46\times 10^{-4}$ & $-2.53 \times 10^{-5}$ & 0 \\
\hline
Latitude $(^\circ\textrm{S})$& 20 or 90 & 10 & 0\\
\hline
$\omega$ $(\mathrm{s}^{-1})$ & $1.4\times 10^{-4}$ or $7.3\times 10^{-5}$ & $-$ & $-$\\
\hline
$U_{\text{mean}}$ $(\si{m.s^{-1}})$& $0$ & $0.2$ & $0.1$\\
\hline
$Fr=Nh_0/U_{\text{mean}}$& $-$ & $0.2$ (2d) or $0.3$ (3d) & $25$\\
\hline
$U_{\text{tidal}}$ $(\si{m.s^{-1}})$& $0.1$ & $0$ & $0$\\
\hline
\end{tabular}
\label{paramtable}
\end{table*}

\subsection{Tidal flow in two dimensions}

\begin{figure*}[h]
    \centering
    \includegraphics[width=0.8\textwidth]{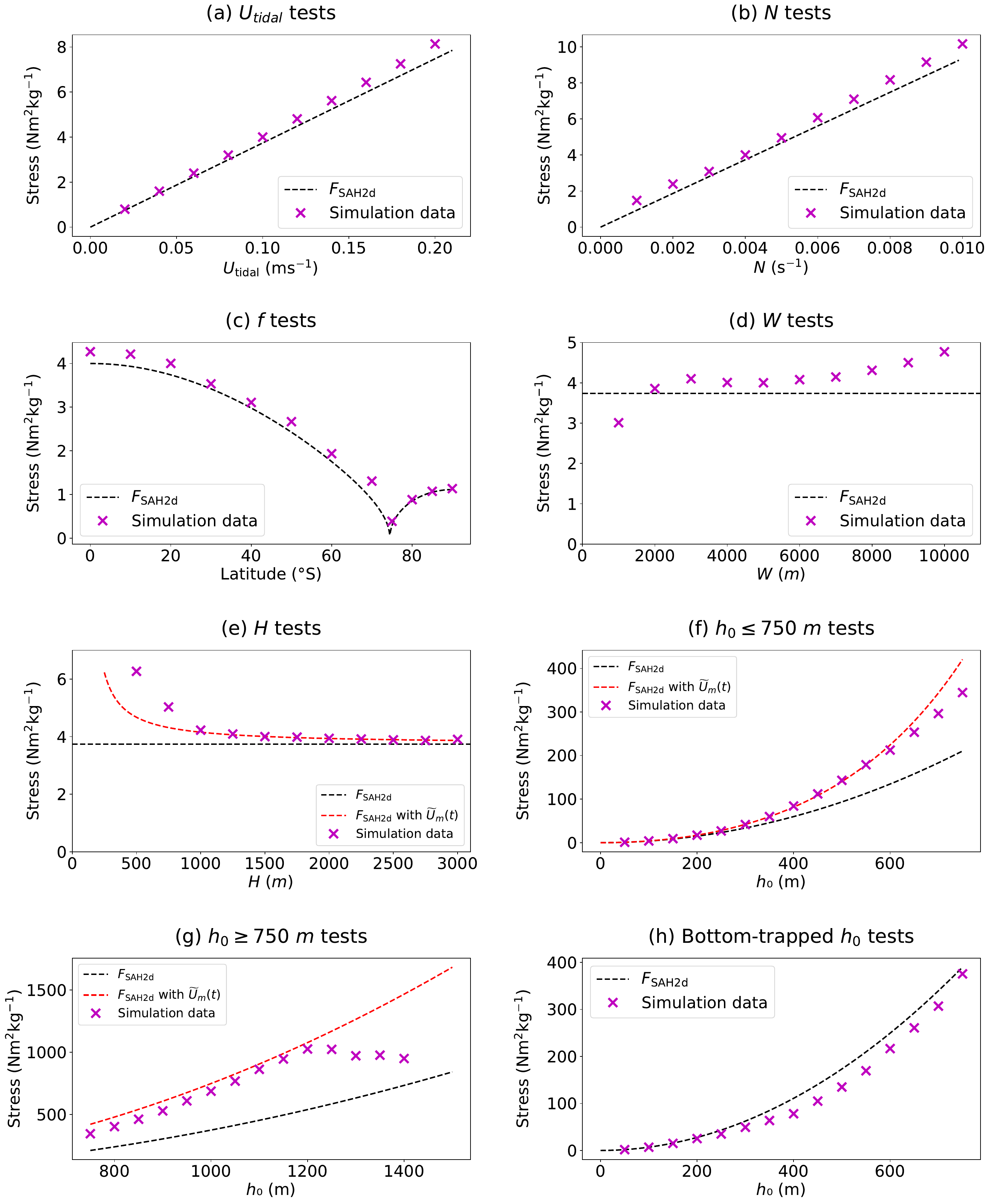}
    \caption{The amplitude of the oscillatory stress given by $F_{\text{SAH2d}}$ and the simulation data for two-dimensional tidal flow. In Figures (e)--(g) a refinement of the $F_{\text{SAH2d}}$ parametrisation with a scaled velocity (see Equation \eqref{tildeum}) is also shown in red. The default parameters for each test are as in Table~\ref{paramtable} with $\omega=10^{-4}\ \si{s^{-1}}$ ($M_2$ tide) at a latitude of $20^\circ S$ ($f = 5 \times 10^{-5}\ \si{s^{-1}}$). The only exception is the bottom-trapped tests in Figure (h) which use $\omega=7.3\times 10^{-5}\ \si{s^{-1}}$ ($K_1$ tide) at a latitude of $90^\circ S$ ($f=1.46\times 10^{-4}\ \si{s^{-1}}$). In all tests, the uncertainty in the computed stress was less than 3\% so the small error bars have been omitted.}\label{fig:2dtidaltests}
\end{figure*}

\begin{figure*}[h]
    \centering
    \includegraphics[width=0.5\textwidth]{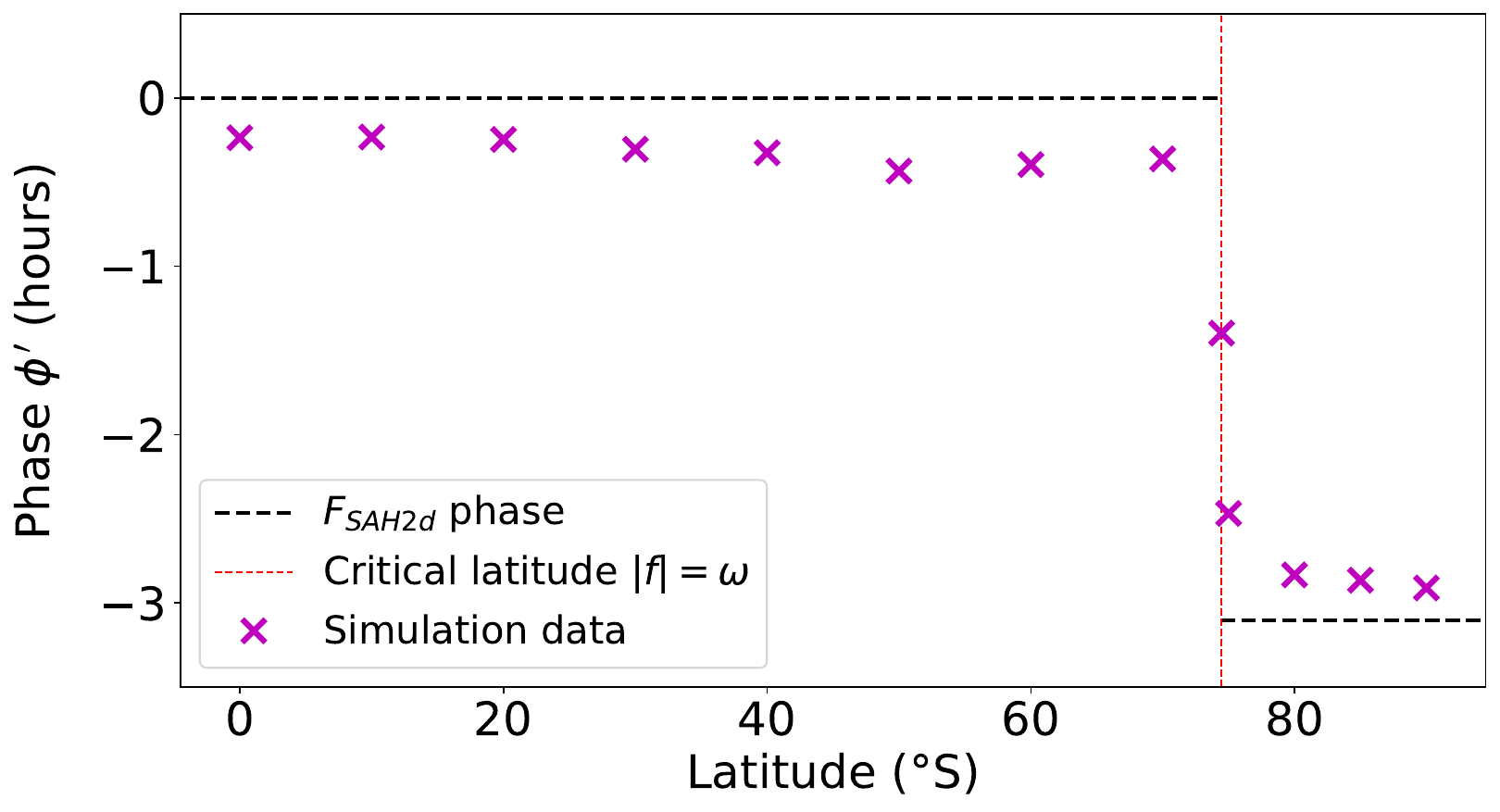}
    \caption{The fitted phase of the stress for our two-dimensional tidal flow simulations varying latitude. Here, the phase $\phi'=\phi\times(\omega/2\pi)$ is given in hours as opposed to radians (cf.\ \eqref{sinesteq}), and the critical latitude ($\approx 74.5^{\circ}S$) is displayed in red. In each test, the uncertainty was less than $0.05$ hours so the small error bars have been omitted.}\label{fig:2dphase}
\end{figure*}

We first consider tidal flow over an isolated hill in two dimensions. For each simulation, the computed stress is compared to the parametrisations in Section~2\ref{secttidalparam}. Figure~\ref{fig:2dtidaltests} shows this comparison for the parametrisation $F_{\text{SAH2d}}$ (see \eqref{SAH2d}) when the parameters $U_{\text{tidal}}$, $N$, $f$, $W$, $H$ and $h_0$ are varied. Although not explicitly shown in Figure~\ref{fig:2dtidaltests}, the simpler JSL2001 parametrisation (Equation \eqref{force-jsl}) predicts the same scaling as $F_{\text{SAH2d}}$ with respect to the stratification $N$, hill height $h_0$ and tidal speed $U_{\text{tidal}}$. Thus our analysis for these variables also holds true for the JSL2001 theory.

For most of the simulations, the flow is forced with the semi-diurnal $M_2$ constituent, with $\omega=1.4\times 10^{-4}\ \si{s}^{-1}$. The only exception is for the bottom-trapped $(|f|>\omega)$ tests displayed in Figure~\ref{fig:2dtidaltests}h, whereby we use the diurnal $K_1$ constituent ($\omega=7.3\times 10^{-5}\ \si{s^{-1}}$). This is because on Earth, bottom-trapped effects are more common for diurnal constituents than semi-diurnal constituents. In particular, for a diurnal constituent the frequency $\omega$ is smaller and thus $|f|>\omega$ for a larger range of latitudes. However, note that we also performed a small number of tests with $|f|>\omega$ for the $M_2$ tide. This appears as part of our tests varying $f$ (Figure~\ref{fig:2dtidaltests}c), in the case when the latitude is greater than $75^{\circ}$.

In general, the parametrisation $F_{\text{SAH2d}}$ performs well compared to the simulations. However, in most cases the theoretical prediction slightly underestimates the stress. This is to be expected as the parametrisations from JSL2001 and SAH2020 are derived in an idealised regime whereby the topographic height and flow speed are assumed to be small. There are thus other non-linear effects that are not accounted for which could contribute to the drag. For instance, calculations of \citet{balmforth2002tidal} show that the energy flux should increase by around 14\% above the linear theory for isolated Gaussian hills as the steepness parameter (defined in \eqref{steepeq}) goes from 0 to 1, which we expect should also translate to form stress.

Looking at Figures~\ref{fig:2dtidaltests}a and \ref{fig:2dtidaltests}b, we see that the predicted linear scaling by the tidal speed $U_{\text{tidal}}$ and the buoyancy frequency $N$ matches the simulations almost precisely. The same is true for the $\sqrt{\omega^2-f^2}/\omega$ scaling for the Coriolis parameter $f$ (Figure~\ref{fig:2dtidaltests}c), highlighting the importance of accounting for latitude in tidal stress, which is neglected in the JSL2001 parametrisation.

However, for the simulations varying the hill width $W$, ocean depth $H$ and hill height $h_0$ (Figures~\ref{fig:2dtidaltests}d--\ref{fig:2dtidaltests}h), there are some larger discrepancies between the computed stress and $F_{\text{SAH2d}}$. Firstly for the tests varying $W$ we see a small increase in the drag as $W$ increases. Note that a priori we do expect some dependence of the stress on $W$ despite it not appearing in $F_{\text{SAH2d}}$. This is because a wider hill gives more area for waves reflected off the surface to interact with topography, as discussed around \eqref{vwaveeq}. Additionally, we also note that the slope criticality (see \eqref{steepeq}) directly depends on the wavenumber $k$ and thus $W$.

Next, for the tests varying the ocean depth $H$ and hill height $h_0$ (Figures~\ref{fig:2dtidaltests}e--g) we find that, provided $|f|<\omega$, the flow in the vicinity of the hill speeds up as it moves over the narrow channel at the top of the hill. As a result, one finds that $F_{\text{SAH2d}}$ underestimates the stress particularly when $h_0$ is large or $H$ is small. Thus, to better model the stress for the $H$ and $h_0$ tests we also consider a refinement of the $F_{\text{SAH2d}}$ parametrisation with a scaled velocity
\begin{equation}\label{uscaling}
    U_m(t)=\frac{H}{H-h_0}U(t),
\end{equation}
inspired by the steady flow parametrisation $F_{KLP}$ (see~\eqref{Umsteadydef}). Here, the factor of $H/(H-h_0)$ is the ratio of the ocean depth $H$ away from the hill, to the ocean depth $H-h_0$ at the centre of the hill. For medium-sized hills replacing $U(t)$ with $U_m(t)$ in $F_{\text{SAH2d}}$ more accurately matches our simulation data. However, we also find that the scaling $U_m(t)$ ``saturates" around $h_0\approx H/2$ in the sense that using the velocity scaling
\begin{equation}\label{tildeum}
    \widetilde{U}_m(t)=
    \begin{cases}
        \dfrac{H}{H-h_0}U(t), &h_0\leq H/2\\
        2\: U(t), &h_0> H/2
    \end{cases}
\end{equation}
yields better results for large values of $h_0$. In particular, for our tests varying $h_0$ in Figures~\ref{fig:2dtidaltests}f and \ref{fig:2dtidaltests}g, we see that incorporating $\widetilde{U}_m(t)$ into the $F_{\text{SAH2d}}$ parametrisation gives a stress profile that is much closer to the simulation data. The only exception is when the hill crest is close to the surface ($h_0\approx H$) and the stress plateaus (Figure~\ref{fig:2dtidaltests}g). We remark that for our tests, there is a discontinuity in the derivative (with respect to $h_0$) of $\widetilde{U}_m(t)$ at $h_0=750\ \si{m}$. However, this is not shown explicitly in Figure~\ref{fig:2dtidaltests} since the discontinuity is at the point which separates Figures \ref{fig:2dtidaltests}f and \ref{fig:2dtidaltests}g. For our tests varying $H$, a similar improvement upon using $\widetilde{U}_m(t)$ can then be seen in Figure~\ref{fig:2dtidaltests}e. However, even with the $\widetilde{U}_m(t)$ scaling, $F_{\text{SAH2d}}$ still underestimates the model output as $H$ decreases. This is not unexpected as wave reflections become more important in this limit (see discussion around  \eqref{vwaveeq}).

We now consider the results for bottom-trapped flows with $|f|\geq\omega$ (Figure~\ref{fig:2dtidaltests}h), which were performed for the K1 tidal constituent at a latitude of $90^{\circ}$. For these tests, in contrast to the $|f|<\omega$ case, we found that the $F_{\text{SAH2d}}$ parametrisation remains relatively accurate as $h_0$ increases (Figure~\ref{fig:2dtidaltests}h). This is expected as such flows have a different velocity profile compared to the $|f|<\omega$ case. Namely, when $|f|\geq\omega$, the flow still speeds up at the top of the hill but is slowed down or ``trapped" at the base of the hill. Thus in this case, it is not accurate to use the $\widetilde{U}_m$ scaling, which assumes the current speeds up along the entire length of the hill.

Finally, as defined in \eqref{sinesteq}, it remains to analyse the frequency $\Omega$ and phase $\phi$ of the simulated stress and how it compares to the parametrisations. Here, we first remark that the fitted frequency from our simulation data closely matches the tidal frequency: in each case the two quantities are within 1\% of each other. However, the phase data is more interesting and for the tests varying $f$, this is shown in Figure~\ref{fig:2dphase}. Here, the $F_{\text{SAH2d}}$ parametrisation suggests that in the absence of wave reflections, the stress should be in phase with $U(t)$ when $|f|<\omega$ and then $90^\circ$ (or 1/4 of a tidal period) out of phase with $U(t)$ when $|f|>\omega$. Note that we give the phase in hours, obtained by multiplying $\phi$ by $\omega/2\pi$. Most of our simulations accurately reflect this prediction, yielding a phase within half an hour of that predicted by $F_{\text{SAH2d}}$. The only exception is when the latitude is very close to the critical value where $|f|=\omega=1.4\times 10^{-4}s$ or approximately $74.5^{\circ}S$. In this exceptional case, the phase is near the half-way point between the predicted subcritical ($|f|<\omega$) and supercritical ($|f|>\omega$) phase. This behaviour is exactly what one would expect from Fourier analysis. However, there would be little practical value in adding this exceptional case to the $F_{\text{SAH2d}}$ parametrisation. Ultimately though, our results for the phase further support including an $f$ dependency in stress parametrisations, which is not present in the JSL2001 parametrisation.

\subsection{Tidal flow in three dimensions}\label{tidalflowthree}
\begin{figure*}[h]
    \centering
    \includegraphics[width=0.8\textwidth]{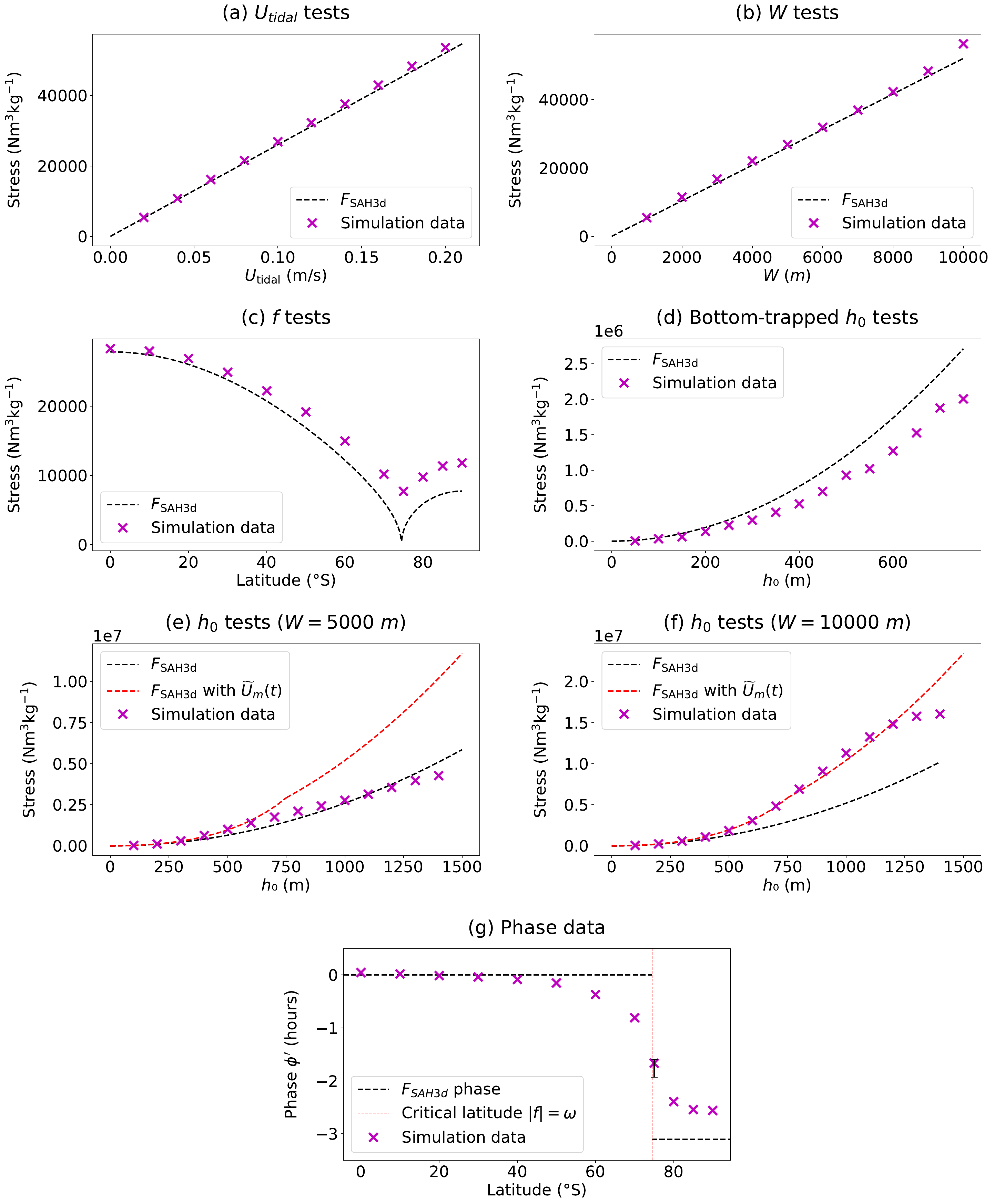}
    \caption{The amplitude of the oscillatory stress given by $F_{\text{SAH3d}}$ and the simulation data for three-dimensional tidal flow (panels~(a)--(f)), and the phase of the stress as the latitude is varied (panel~(g)). The default values for each test are the same as the two-dimensional case (see Figure~\ref{fig:2dtidaltests} and Table~\ref{paramtable}). In panels~(a)--(f), the uncertainty in the computed values was less than 10\% so the small error bars have been omitted. In Figure (g), error bars have only been included in the single case where the uncertainty in the phase was greater than 0.1 hours.}\label{fig:3dtidaltests}
\end{figure*}
Our results for tidal flow in three dimensions are analysed analogously to our results in two dimensions. Note that in what follows, we compute the $x$-component of the stress amplitude, frequency and phase. Figure~\ref{fig:3dtidaltests} shows a comparison between $F_{\text{SAH3d}}$ (Equation \eqref{SAH3d}) and the simulation data as the tidal speed $U_{\text{tidal}}$, hill width $W$, Coriolis parameter $f$ and hill height $h_0$ are varied. Plots of the phase are also included for tests that vary the latitude (or equivalently $f$).

Similar to the two-dimensional case, the simulations varying $U_{\text{tidal}}$, $W$ and $f$ (Figures~\ref{fig:3dtidaltests}a--\ref{fig:3dtidaltests}c) qualitatively reflect the scaling predicted by the $F_{\text{SAH3d}}$ parametrisation.

For the tests varying $f$, the $F_{\text{SAH3d}}$ parametrisation well approximates the model output for near-zero $|f|$. However, there is an almost linearly increasing error as $f$ increases. As this effect did not occur in two-dimensions, there must be something specific to three-dimensions causing this discrepancy. As $|f|$ increases, there is increasing flow in the $y$  direction which in three-dimensions (unlike two) is permitted to vary in $y$, and we speculate that this additional degree of freedom is likely responsible for the growing error, although the details remain uncertain.

We now consider tests varying $h_0$. In the bottom-trapped ($|f|>\omega$) case, the simulated data matches the scaling predicted by the $F_{\text{SAH3d}}$ parametrisation (Figure~\ref{fig:3dtidaltests}d). However, for our choice of parameters, $F_{\text{SAH3d}}$ overestimates the simulated stress by more than in the two-dimensional case. In particular, over all the tests we performed, $F_{\text{SAH3d}}$ was on average $46\%$ larger than the simulated stress in three dimensions, compared to $18\%$ for $F_{\text{SAH2d}}$ in two-dimensions. In the case where $|f|<\omega$ there is then a qualitative difference between the computed stress profile and that predicted by $F_{\text{SAH3d}}$, depending on whether the hill width $W$ is small or large (Figures~\ref{fig:3dtidaltests}e and \ref{fig:3dtidaltests}f). To account for this difference, we let $\widetilde{F}_{\text{SAH3d}}$ be equal to $F_{\text{SAH3d}}$ but with the velocity scaling $\widetilde{U}_m$ (Equation \eqref{tildeum}) discussed in the two-dimensional setting. For large $W$ (Figure~\ref{fig:3dtidaltests}f), the scaling of the stress is the same as in the two-dimensional case. Namely, $\widetilde{F}_{\text{SAH3d}}$ matches the stress profile until $h_0\approx H$. However, for our default width $W=5000\ m$, $\widetilde{F}_{\text{SAH3d}}$ only matches the simulated stress for small--medium values of $h_0$ (Figure~\ref{fig:3dtidaltests}e). Then, for $h_0\gtrsim 500\ m$, the stress is much closer to the unscaled $F_{\text{SAH3d}}$ parametrisation. Since $\widetilde{F}_{\text{SAH3d}}$ is only slightly larger than $F_{\text{SAH3d}}$ for small hill heights, it thus seems reasonable to just apply the unscaled parametrisation $F_{\text{SAH3d}}$ for all values of $h_0$ in this ``small $W$" regime. We remark that this regime shift for smaller $W$ is not entirely unexpected, as provided $W$ is not too large, we expect some of the current to flow around the hill, rather than be forced over the top as in the two-dimensional case. Future work should identify more precisely at what values of $W$ this regime shift occurs.

Finally, we remark that, as with the two-dimensional case, the frequency and phase accurately matches the values predicted by the SAH2020 theory. However, the most notable difference in three dimensions is a smoother transition from the stress being in phase when $|f|<\omega$ to being $90^\circ$ out of phase when $|f|>\omega$ (Figure~\ref{fig:3dtidaltests}h). Nevertheless, this effect was only clear when $|f|$ was very close to $\omega$ so that the phase predicted in $F_{\text{SAH3d}}$ is still accurate for almost all values of $f$.

\subsection{Steady flow in two dimensions}

\begin{figure*}[h]
    \centering
    \includegraphics[width=0.9\textwidth]{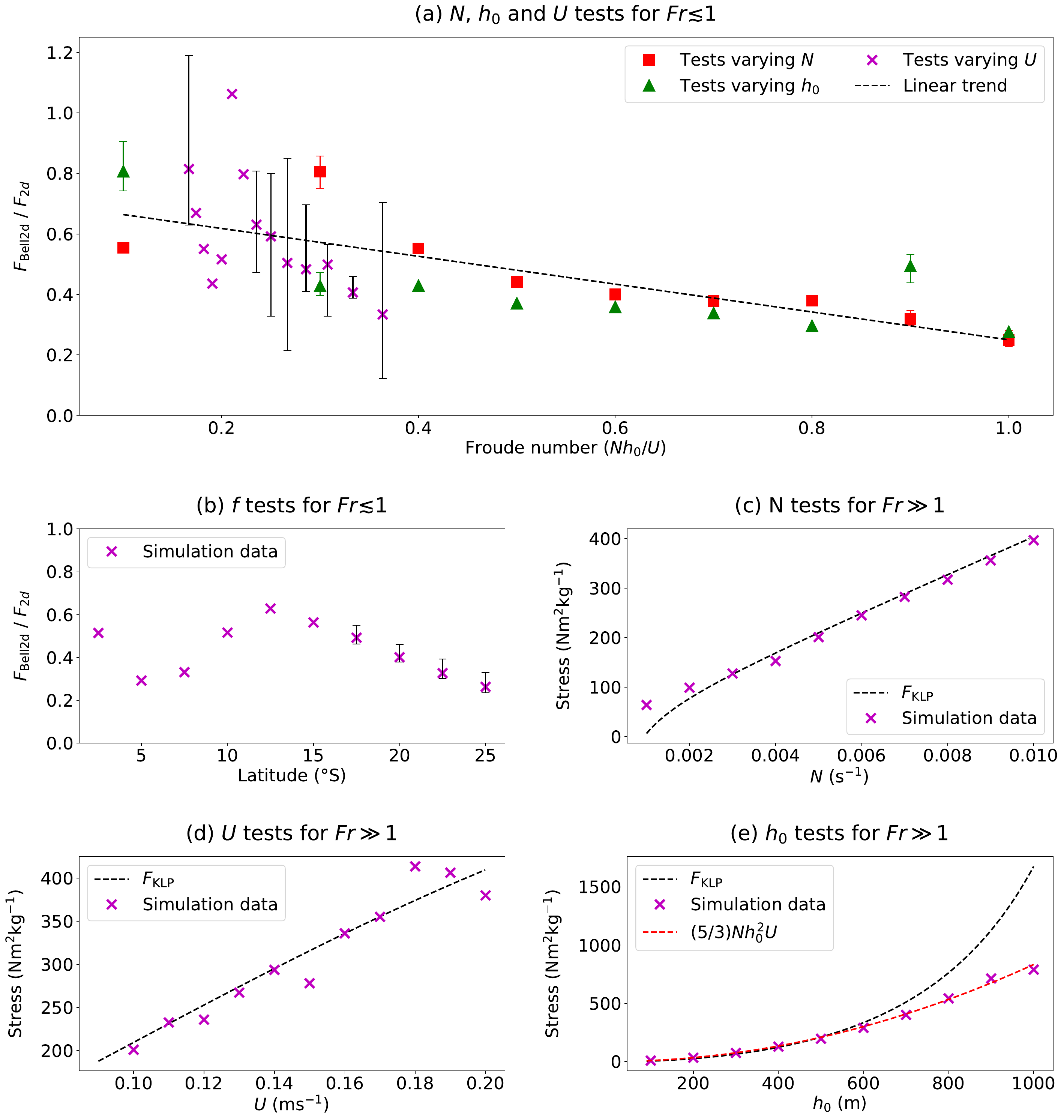}
    \caption{A comparison of the stress predicted by the \citet{bell1975topographically} and KLP2010 parametrisations with our simulation data for two-dimensional steady flow. Here, $N$ denotes the buoyancy frequency, $h_0$ the hill height, $U$ the flow speed, and $f$ the Coriolis parameter. Note that for Figures (a) and (b), the ratio between the stress as predicted the \citet{bell1975topographically} theory, $F_{\text{Bell2d}}$, and the stress output from the simulations, $F_{\text{2d}}$,  is plotted. On the other hand, in Figures (c) to (e) the numerical values of the stress predicted by KLP2010 parametrisation, $F_{\text{KLP}}$, is shown. The default parameters are as in Table~\ref{paramtable} and error bars are shown whenever the uncertainty in the stress was greater than 10\%.}\label{fig:2dsteadytests}
\end{figure*}

For two-dimensional tests with a steady flow we compare the computed stress (denoted $F_{\text{2d}}$) with the parametrisations $F_{\text{Bell2d}}$ and $F_{\text{KLP}}$ from Section~\ref{existingsect}. In particular, tests with low Froude numbers $Fr\lesssim 1$ are compared with $F_{\text{Bell2d}}$ and tests with high Froude numbers $Fr\gg 1$ are compared with~$F_{\text{KLP}}$.

Figure~\ref{fig:2dsteadytests}a shows our comparison between the simulated data and $F_{\text{Bell2d}}$ for Froude numbers in the range $0.1-1$. We see that although the scaling used in Bell's theory is generally accurate, $F_{\text{Bell2d}}$ typically underestimates the stress by about $50\%$, and even more so as the Froude number approaches 1. Hence, Bell's parametrisation should ideally be adjusted to depend on $Fr$. In this regard, one could define a scaled parametrisation
\begin{equation}\label{tildebell2d}
    \widetilde{F}_{\text{Bell2d}}=\frac{1}{A_1-B_1(Nh_0/U)}F_{\text{Bell2d}},
\end{equation}
for constants $A_1$, $B_1$, and $Fr\lesssim 1$. Note that the linear trend line in Figure~\ref{fig:2dsteadytests}a corresponds to the values $A_1=0.71$ and $B_1=0.46$. However, a parametrisation of the form \eqref{tildebell2d} still does not describe the full behaviour of the stress in the regime $Fr\lesssim 1$. Namely, in Figure~\ref{fig:2dsteadytests}b we also see a non-linear relationship between the stress and latitude. This reinforces that, even for low Froude numbers, non-wave and non-linear effects manifest that are not accounted for by Bell's theory. We also remark that the trend line in Figure~\ref{fig:2dsteadytests}a is the opposite to what one gets with energy flux calculations. That is, as the Froude number increases, one reaches a saturation of lee waves and Bell's formula for energy flux becomes an overestimate (see \eqref{frcriteq}).

Figures~\ref{fig:2dsteadytests}c--\ref{fig:2dsteadytests}e then show our comparison between the simulated data and $F_{\text{KLP}}$ for large Froude numbers and $f=0$. Here, the approximately linear scaling of buoyancy frequency $N$ and flow speed $U$ in $F_{\text{KLP}}$ seems to be very accurate (Figures~\ref{fig:2dsteadytests}c and \ref{fig:2dsteadytests}d). On the other hand, the height scaling of $h_0^2 H/(H-h_0)$ overestimates the stress as the hill height $h_0$ gets large (Figure~\ref{fig:2dsteadytests}e). Our results indicate that a simpler scaling than that in $F_{\text{KLP}}$ is more accurate. In particular, if we parametrise the stress as
\begin{equation}\label{newkly}
    \text{Stress}=C_1 Nh_0^2 U,
\end{equation}
for some suitable constant $C$, then we get much closer to the drag from the simulations. For $C_1=1.4$ this is shown in Figure~\ref{fig:2dsteadytests}e. Moreover, using \eqref{newkly} as opposed to $F_{\text{KLP}}$ gives a valid result for lower Froude numbers when $F_{\text{KLP}}$ is negative (and thus unphysical).

We also ran a few tests with rotation added to the high Froude number simulations. In this case, the stress was unstable and increased by several fold. Upon analysing the simulation data, we found that for these cases significant mixing occurred downslope, and gradually reduced the downstream stratification to zero. This phenomena is reminiscent of atmospheric downslope windstorms, which can occur near large-scale topography on land \citep[see e.g.,][]{peltier1979evolution,scinocca2000parametrization}. However, this effect was far less dramatic in our three-dimensional simulations, where the presence of a lateral flow helps to stabilise the stratification. Such effects are thus likely to be rare in practice, mostly restricted to ridge-like structures. Rather than attempting to parametrise this non-linear phenomena here, we instead refer the reader to the recent work of \citet{zemskova2022energetics}. Here, the energetics of a steady rotating flow over tall topography is analysed in detail, and used to help explain observations in the Southern Ocean by \citet{cusack2020observed}.

\subsection{Steady flow in three dimensions}\label{steady3results}
\begin{figure*}[h]
    \centering
    \includegraphics[width=0.9\textwidth]{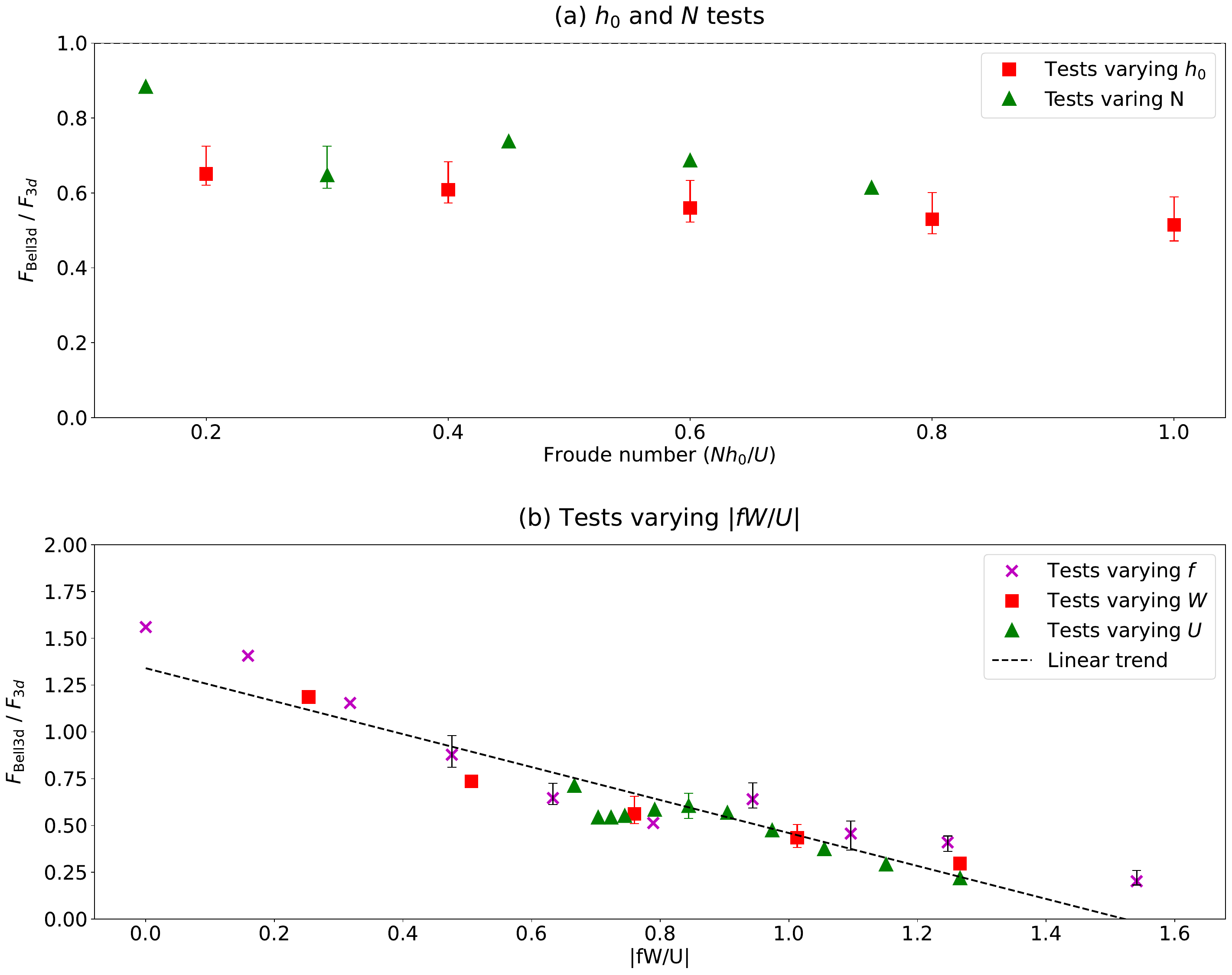}
    \caption{A comparison of the stress predicted by the \citet{bell1975topographically} parametrisation ($F_{\text{Bell3d}}$) and the simulation drag ($F_{3d}$) for three-dimensional steady flows with low Froude number $(Fr\lesssim 1)$. The default parameters are as in Table~\ref{paramtable} and error bars are shown whenever the uncertainty in the computed stress is greater than 10\%.}\label{fig:3dsteadytestslowF}
\end{figure*}

\begin{figure*}[h]
    \centering
    \includegraphics[width=0.9\textwidth]{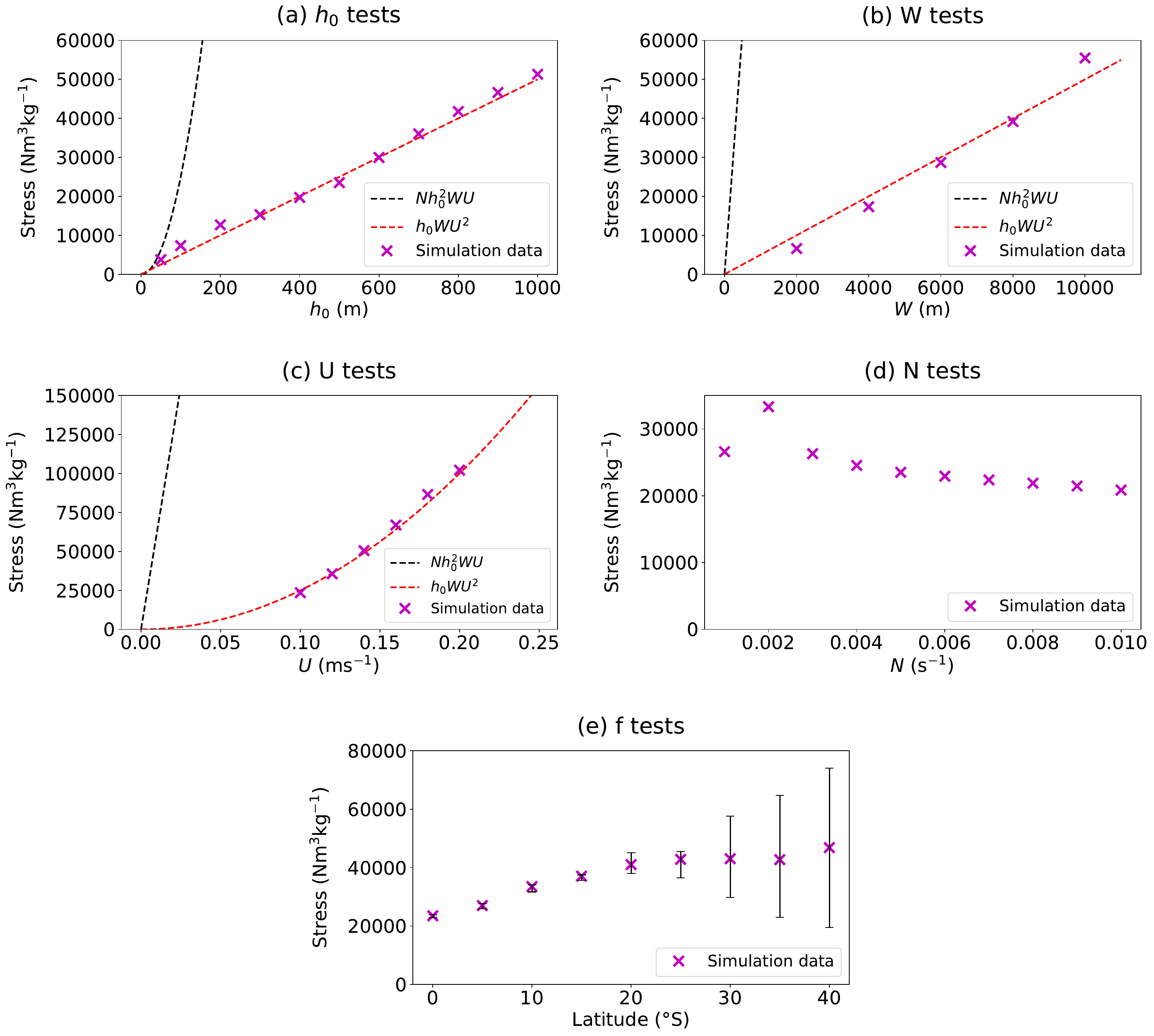}
    \caption{Topographic stress for three-dimensional steady flows with high Froude number $(Fr\gg 1)$. In panels~(a)--(c), the drag from the simulation data is compared with both a linear $Nh_0^2WU$ and quadratic $h_0WU^2$ scaling. The default parameters are as in Table~\ref{paramtable}. Note that in panels~(a)--(d) the uncertainty in the computed stress was always less than 10\% so the small error bars have been omitted.}
    \label{fig:3dsteadytestshighF}
\end{figure*}

For steady flow tests in three-dimensions, we again compare simulations with low Froude numbers $Fr\lesssim 1$ to the \citet{bell1975topographically} theory. However, for tests with high Froude numbers $Fr\gg 1$, it does not make sense to compare our results directly to KLP2010 as this theory was only derived in the two-dimensional case. Thus, for high Froude numbers, we instead focus on how the stress scales with different parameters.

Figure~\ref{fig:3dsteadytestslowF} shows a comparison between the stress from our simulations (denoted $F_{\text{3d}}$) and $F_{\text{Bell3d}}$ when the parameters $h_0$, $N$, $U$, $f$ and $W$ are varied. For tests varying the hill height $h_0$ and buoyancy frequency $N$ (Figure~\ref{fig:3dsteadytestslowF}a) we find, similar to the two-dimensional case, that $F_{\text{Bell3d}}$ underestimates the stress but still gives a mostly accurate scaling (i.e.\ $\text{Stress}\propto Nh_0^2$). However, for tests that vary the flow speed $U$, Coriolis parameter $f$ and hill width $W$ (Figure~\ref{fig:3dsteadytestslowF}b), we see that $F_{\text{Bell3d}}$ significantly underestimates the stress as the Rossby number
\begin{equation}
    Ro \equiv \frac{U}{|f|W},
\end{equation}
decreases. In this regard, it is worth noting that via a change of variables, $F_{\text{Bell3d}}$ can be expressed as
\begin{align}\label{polarbell}
    &F_{\text{Bell3d}}\left(\alpha=0,\ h(x,y)=h_0e^{-(x^2+y^2)/2W^2}\right) = \notag\\
    &\ =\mathrm{Re}\left(\int\limits_0^{2\pi}\!\!\!\!\int\limits_0^\infty \!\! \tilde{K} e^{-\tilde{K}^2}\cos^2\theta\sqrt{\tilde{K}^2-\frac{1}{Ro^2\cos^2\theta}}\,\mathrm{d}\tilde{K}\mathrm{d}\theta\right)Nh_0^2 W\rho_0
\end{align}
where $\tilde{K}=W\sqrt{k^2+\ell^2}$ and $(kW,\ell W)=(\tilde{K}\cos\theta, \tilde{K}\sin\theta)$. From \eqref{polarbell} we see that as $Ro$ increases so too does the lee wave stress. Therefore, a possible explanation for the discrepancy between $F_{\text{Bell3d}}$ and the simulation data is that as $Ro$ decreases, fewer lee waves are formed and the stress is instead generated by other non-linear motions such as partial blocking \citep[see e.g.,][]{trossman2016impact,klymak2021parameterizing}. To parametrise this phenomenon, we note that from Figure~\ref{fig:3dsteadytestslowF}b the ratio $F_{\text{Bell3d}}/F_{3d}$ has a clear linear dependence on the ratio $|f|W/U$. In particular, our data points give a correlation coefficient between $F_{\text{Bell3d}}/F_{3d}$ and $|f|W/U$ of $R^2=0.89$. This suggests that we should use the scaling
\begin{equation}\label{fwUbelleq}
    \widetilde{F}_{\text{Bell3d}}=\frac{1}{A_2-B_2|fW/U|}F_{\text{Bell3d}}
\end{equation}
for some constants $A_2$ and $B_2$ and $Fr\lesssim 1$. The linear trend line in \ref{fig:3dsteadytestslowF}b corresponds to $A_2=1.34$ and $B_2=0.88$. Certainly, \eqref{fwUbelleq} should only be applied for small values of $|fW/U|$ so that the denominator $A_2-B_2|fW/U|$ remains positive.

Figure~\ref{fig:3dsteadytestshighF} shows the results of our tests for the high Froude number $Fr\gg 1$ regime. In this case, we find that the stress deviates significantly from the approximate $Nh_0^2 W U$ scaling found in other regimes. Instead we find that the stress obeys a quadratic law, being proportional to $h_0WU^2$, similar to what one expects in the atmosphere \citep[e.g.,][]{lott1997new}. In fact, the simple parametrisation, which sets the stress equal to $h_0WU^2$ (with proportionality constant $1$) works remarkably well for our choice of parameters. However, as is usual with quadratic drag laws, it is likely that a different proportionality constant would be required if the shape of the topography differed substantially from a Gaussian hill. The stress then appears to be mostly independent of $N$ and $f$ (Figures~\ref{fig:3dsteadytestshighF}d and \ref{fig:3dsteadytestshighF}e); only increasing or decreasing by small amounts as these parameters are varied. We note however, that for larger values of $f$, our computed value of the stress has a large uncertainty. These results can be directly compared to recent work of \citet{klymak2021parameterizing} which involved a similar analysis but with many, randomly distributed obstacles. Klymak et al.\ also observed this $f$ independence for the stress but an approximately linear dependence on $N$ for high Froude numbers. Therefore, we do not expect the stress to remain independent of $N$ for more complicated non-isolated topography.

\subsection{Mixed flow in two dimensions}\label{mixedtworesults}
\begin{figure*}[h]
    \centering
    \includegraphics[width=0.75\textwidth]{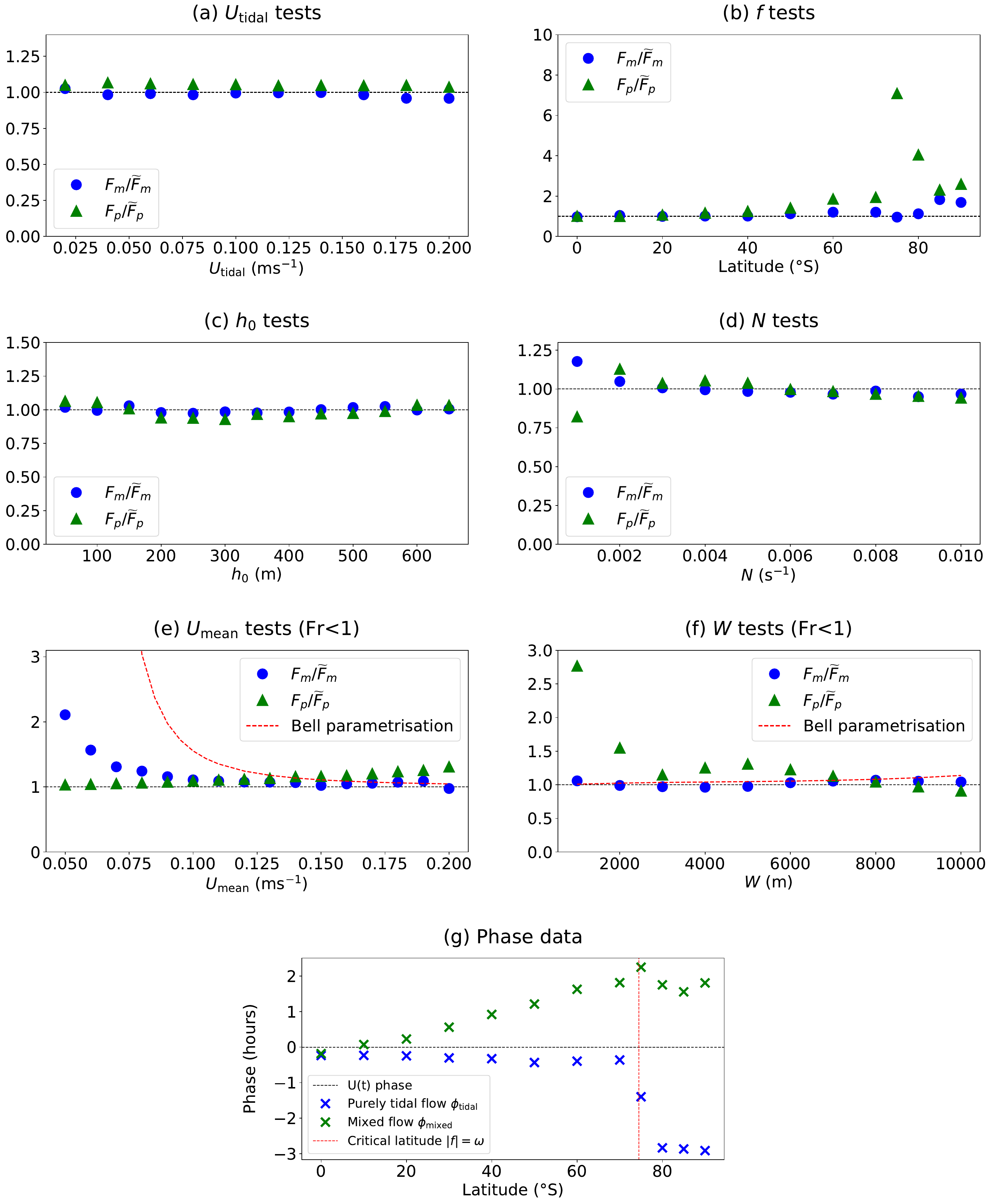}
    \caption{Results for two-dimensional mixed flow simulations. In panels~(a)--(d), the value of the ratios $F_m/\widetilde{F}_m$ and $F_p/\widetilde{F}_p$ are shown using $U_{\text{mean}}=0.1$ $\mathrm{ms}^{-1}$ and the default parameters from our tidal tests (Figure~\ref{fig:2dtidaltests}). In panels~(e) and~(f), our results for low Froude numbers are displayed using $U_{\text{tidal}}=0.1$ $\mathrm{ms}^{-1}$, an $M_2$ tidal frequency, and the default parameters from our steady flow tests (Figure~\ref{fig:2dsteadytests}). In Figure (g), our results for the phase shift for both the tidal ($\widetilde{F}_p$) and mixed ($F_p$) cases are shown for the simulations that varied latitude (cf.\ Figure (b)). In each test, the uncertainty was always less than 10\% (or less than 0.2 hours for Figure (g)) so the small error bars have been omitted.}\label{fig:2d_mixed_plots}
\end{figure*}

We now analyse our results for mixed flows with a horizontal velocity of the form \eqref{Uequation} and stress of the form \eqref{sinestmix}. In what follows we let $\widetilde{F}_m$ and $\widetilde{F}_p$ be the values for $F_m$ and $F_p$ if $U_{\text{tidal}}$ or $U_{\text{mean}}$ are set to be zero respectively. In Figure~\ref{fig:2d_mixed_plots} we show our computed values of $F_m/\widetilde{F}_m$ and $F_p/\widetilde{F}_p$ for a range of parameters. This lets us discern the impact of a tidal flow on a steady flow and vice versa. When the Froude number is small ($Fr\lesssim 1$) these ratios are also compared to the expected ratio from Bell's parametrisations defined in Sections 2\ref{steadytheorysect} and 2\ref{mixedtheorysect} (Figures~\ref{fig:2d_mixed_plots}e and \ref{fig:2d_mixed_plots}f).

To compute the uncertainty in our values of $F_m/\widetilde{F}_m$ and $F_p/\widetilde{F}_p$, we define $F_m^-$, $\widetilde{F}_m^-$, $F_p^-$, $\widetilde{F}_p^-$ and $F_m^+$, $\widetilde{F}_m^+$, $F_p^+$, $\widetilde{F}_p^+$ to respectively be the lower and upper values of uncertainty for each quantity. The lower and upper uncertainties in $F_m/\widetilde{F}_m$ and $F_p/\widetilde{F}_p$ are then given by
\begin{equation}
    \left(\frac{F_{m}}{\widetilde{F}_m}\right)^{\pm}=\frac{F_m^{\pm}}{\widetilde{F}_m^{\mp}}\qquad\text{and}\qquad\left(\frac{F_p}{\widetilde{F}_p}\right)^{\pm}=\frac{F_p^{\pm}}{\widetilde{F}_p^{\mp}}.
\end{equation}

First we consider our results for the mean component of the stress $F_m$. In general, for both low and high Froude numbers, we find that $F_m/\widetilde{F}_m\approx 1$, suggesting that the naive parametrisation $F_m=\widetilde{F}_m$ is accurate for a wide range of parameters. However, in some regimes we find that $F_m$ notably exceeds $\widetilde{F}_m$. This is most evident in Figure~\ref{fig:2d_mixed_plots}e, where $F_m$ can be over two times larger than $\widetilde{F}_m$ for small values of $U_{\text{mean}}$. This is reflected by Bell's formula \eqref{bellmixed2d}, whereby as $U_{\text{mean}}\to 0$ we have $\omega_0\to 0$ so that the $n=0$ term becomes small and the $n\geq 1$ terms associated with the tide become more dominant. However, we note that using Bell's parametrisation, $F_{m}$ exceeds $\widetilde{F}_m$ by more than what is reflected by the simulation data.

Our simulations also indicate that Bell's parametrisation is very inaccurate for larger Froude numbers $Fr\geq 1$, so that this parametrisation should not be extended to such a case. For instance, for our $h_0$ tests in Figure~\ref{fig:2d_mixed_plots}c we have ${2\leq Fr\leq 26}$ and $F_{m}/\widetilde{F}_m\approx 1$. However, Bell's parametrisation gives the substantially larger value $F_{m}/\widetilde{F_m}=21.38$.

Next we consider our results for the periodic component of the stress $F_p$. Similar to the mean component, we find in most regimes that $F_p/\widetilde{F}_p\approx 1$, thereby further highlighting the accuracy of the naive parametrisation $F_p=\widetilde{F}_p$. The most notable exception is near critical latitudes when $|f|\approx\omega$ (Figure~\ref{fig:2d_mixed_plots}b). This is to be expected as the steady component of the flow acts to Doppler shift the tidal frequency $\omega$. Therefore when $|f|=\omega$, we expect $F_p$ to not approach zero unlike $\widetilde{F}_p$ (see \eqref{SAH2d}), leading to a large value of the ratio $F_p/\widetilde{F}_p$.

For low Froude numbers (Figures~\ref{fig:2d_mixed_plots}e and \ref{fig:2d_mixed_plots}f) we also observe a dependence of $F_p/\widetilde{F}_p$ on the mean flow speed $U_{\text{mean}}$ and the hill width $W$. In particular, $F_p/\widetilde{F}_p$ slightly exceeds 1 as $U_{\text{mean}}$ increases (Figure~\ref{fig:2d_mixed_plots}e), and significantly exceeds $1$ for small $W$ (Figure~\ref{fig:2d_mixed_plots}f). This is likely linked to the fact that the classical lee wave frequency $kU_{\text{mean}}$ depends explicitly on $U_{\text{mean}}$ and implicitly on $W$. In particular, with the Fourier convention in~\eqref{eq:Fourierconvention}, the wavenumber spectrum for a Gaussian hill $h(x)=h_0e^{-x^2/(2W^2)}$ is
\begin{equation}
    \hat{h}(k)= h_0W\sqrt{2\pi}e^{-k^2W^2/2}.
\end{equation}
In terms of developing a refined parametrisation, one could fit a linear trend line for $F_p/\widetilde{F}_p$ against $U_{\text{mean}}$. However, we are not sure how useful this would be since the increase in $F_p/\widetilde{F}_p$ is only small with $U_{\text{mean}}$ and it is clear that $W$ is the far more sensitive parameter. It appears difficult however, to develop a more accurate parametrisation for small $W$, especially given how non-linear the trend is between $F_p/\widetilde{F}_p$ and $W$ in Figure \ref{fig:2d_mixed_plots}f. Notably, low values of $W$ correspond to large values of the steepness parameter (see \eqref{steepeq}), so that non-linear effects are expected to take place in this regime.

We also investigate the change in phase from a purely tidal flow to a mixed flow. In what follows we set $\phi_{\text{tidal}}$ and $\phi_{\text{mixed}}$ to be the respective phases for a purely tidal and a mixed flow with the same choice of parameters (besides the presence of a mean flow $U_{\text{mean}}=0.1$ $\si{m.s^{-1}}$ in the mixed case). When $|f|\ll\omega$, we find that both $\phi_{\text{tidal}}$ and $\phi_{\text{mixed}}$ are approximately in phase with the velocity~$U(t)$. However, as $f$ increases, $\phi_{\text{tidal}}$ increasingly deviates from $\phi_{\text{mixed}}$. In Figure~\ref{fig:2d_mixed_plots}g this is shown for our tests varying latitude (cf.\ Figure~\ref{fig:2d_mixed_plots}b), whereby the difference between $\phi_{\text{tidal}}$ and $\phi_{\text{mixed}}$ reaches about 5 hours at $80^\circ S$. This effect was also observed in a small number of supplementary simulations (not shown) for other tidal constituents besides $M_2$.

\subsection{Mixed flow in three dimensions}\label{mixedthreeresults}

\begin{figure*}[h]
    \centering
    \includegraphics[width=0.75\textwidth]{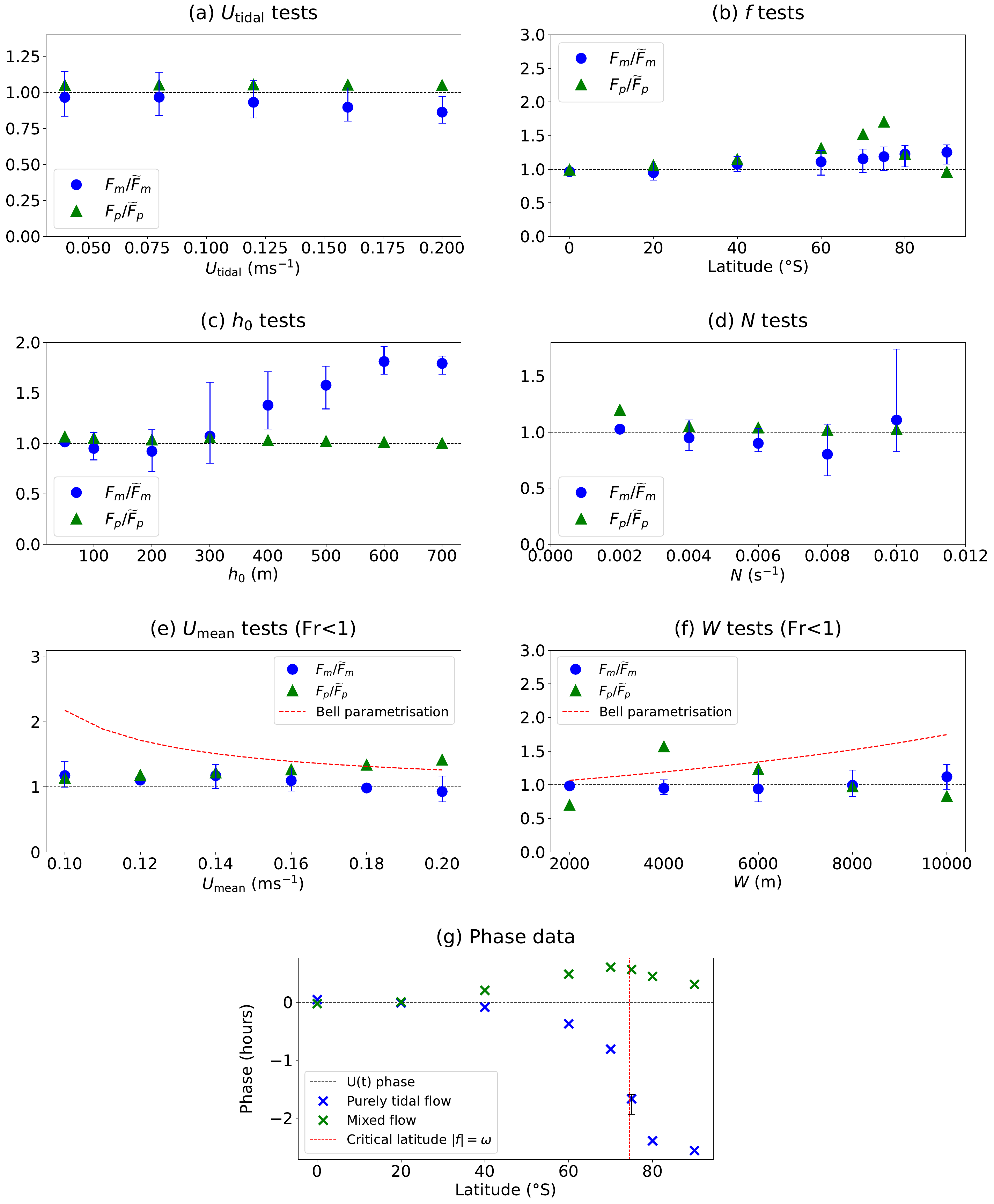}
    \caption{Results for three-dimensional mixed flow simulations. In panels~(a)--(d), the value of the ratios $F_m/\widetilde{F}_m$ and $F_p/\widetilde{F}_p$ are shown using $U_{\text{mean}}=0.1$ $\mathrm{ms}^{-1}$ and the default parameters from our tidal tests (Figure~\ref{fig:3dtidaltests}). In panels~(e) and~(f), our results for low Froude numbers are displayed using $U_{\text{tidal}}=0.1$ $\mathrm{ms}^{-1}$, an $M_2$ tidal frequency, and the default parameters from our steady flow tests (Figure~\ref{fig:3dsteadytestslowF}). In Figure (g), our results for the phase shift for both the tidal ($\widetilde{F}_p$) and mixed ($F_p$) cases are shown for the simulations that varied latitude (cf.\ Figure (b)). In panels~(a)--(f), error bars are only shown when the uncertainty is greater than 10\%. In Figure (g), error bars are only displayed in the single case where the uncertainty was greater than 0.2 hours.}\label{fig:3d_mixed_plots}
\end{figure*}

Finally we consider mixed flows in three dimensions, with our results for this regime shown in Figure~\ref{fig:3d_mixed_plots}. Note that we use the same notation as in Section~4\ref{mixedtworesults}). Like the two-dimensional case, we also find that $F_m/\widetilde{F}_m\approx 1$ and $F_p/\widetilde{F}_p\approx 1$ for the majority of tests. In fact, this approximation appears to hold true even more than in the two-dimensional case (cf. Figure~\ref{fig:2d_mixed_plots}).

Of particular note is the far less dramatic increase in $F_p/\widetilde{F}_p$ near the critical latitude $|f|\approx\omega$. This behaviour seems primarily related to our previous observations for purely tidal flows, rather than some new phenomena specifically linked to three-dimensional mixed flows. Namely, in two-dimensions $\widetilde{F}_p$ essentially reaches 0 at the critical latitude (Figure~\ref{fig:2dtidaltests}c) causing the periodic component of the mixed (and thus Doppler-shifted) stress $F_p$ to be very large by comparison. Then, by contrast, in three dimensions the value of $\widetilde{F}_p$ does not approach zero when $|f|=\omega$ (Figure~\ref{fig:3dtidaltests}c), resulting in a more subdued peak for the ratio $F_p/\widetilde{F}_p$.

Another notable observation in Figure~\ref{fig:3d_mixed_plots} is
the distinct increase in the ratio $F_m/\widetilde{F}_m$ as $h_0$ increases (Figure~\ref{fig:3d_mixed_plots}c), which was not observed in the two-dimensional setting. To try to understand this increase, we recall the results from our purely steady flow tests with large Froude number in three dimensions (see Figure~\ref{fig:3dsteadytestshighF}). For this regime, the mean stress followed a quadratic drag law, being proportional to $h_0 WU^2$ as opposed to the $h_0^2 WU$ scaling for low Froude numbers. Thus, averaging over one tidal period $T$, one expects that for large Froude numbers
\begin{align}
    \frac{F_m}{\widetilde{F_m}}& = \frac{1}{h_0WU_{\text{mean}}^2T} \int_0^{T}h_0W\left[U_{\text{mean}} + U_{\text{tidal}}\cos(\omega t)\right]^2\mathrm{d}t\\
    & = 1 + \frac{U_{\text{tidal}}^2}{2U_{\text{mean}}^2}.\label{largeFFmeq}
\end{align}
For the case $U_{\text{tidal}}=U_{\text{mean}}=0.1$ represented in Figure~\ref{fig:3d_mixed_plots}c, this gives $F_m/\widetilde{F}_m=1.5$, which is not too far off the model output for the larger values of $h_0$. In future work, it would be useful to understand this phenomenon further. In particular, it would be worthwhile evaluating the formula \eqref{largeFFmeq} throughout a wider parameter space. Moreover, it would also be useful to better understand the location and behavior of the transition between the low and high Froude number regime. Based on the limited data in Figure \ref{fig:3dtidaltests} this transition appears around $Fr\approx 10-20$, and corresponds to larger uncertainty in $F_m/\widetilde{F}_m$.

The other key difference in the three-dimensional setting is the value of the phase for mixed flows. In Figure~\ref{fig:3d_mixed_plots}g we see that the mixed flow phase is close to 0 so that the total stress is given by
\begin{equation}
    F_{3d}=F_m+F_p\cos(\omega t+\phi),
\end{equation}
with $\phi\approx 0$. By comparison, in two dimensions we found that $\phi$ was much larger as $|f|$ increased (Figure~\ref{fig:2d_mixed_plots}g). Such an effect is also in contrast to the purely tidal flow case, whereby the phase $\phi$ approaches $-\pi/2$ for large $|f|$.

\section{Discussion and summary of suggested parametrisations}\label{summsect}
In the previous section, we tested existing topographic stress parametrisations for oceanic flows against a suite of simulations. Particular parametrisations of interest were those of \citet{jayne2001parameterizing} and \citet{shakespeare2020drag} for tidal flows, as well as \citet{bell1975topographically} and \citet{klymak2010high} for steady flows. Spanning a wide parameter space for an isolated Gaussian hill, we found that in many cases these parametrisations proved accurate, in many cases even when tested beyond the validity of their underlying assumptions. However, there were also regimes in which these existing parametrisations fell short, suggesting the need for an alternate or modified parametrisation.

We also studied the case of a mixed flow in Sections~4\ref{mixedtworesults} and 4\ref{mixedthreeresults}. Here, we primarily compared our simulation data against the naive parametrisation which assumes that the tidal and steady components of a flow are independent. Our data yielded the cases for which this naive parametrisation was accurate, and cases for which a more sophisticated parametrisation needs to be developed.

In what follows, we summarise our suggested parametrisations based on our results in Section~\ref{resultssect}. Here, as in Section~\ref{existingsect}, $F_{\text{2d}}$ and $F_{\text{3d}}$ denote the stress in two and three dimensions respectively.

\subsection{Suggested tidal flow parametrisations}
For a purely tidal flow, we found that the parametrisations $F_{\text{SAH2d}}$ (Equation~\eqref{SAH2d}) and $F_{\text{SAH3d}}$ (Equation~\eqref{SAH3d} due to \citet{shakespeare2020drag,shakespeare2021impact} were very accurate, mainly just requiring a small modification for large hill heights. In particular, for a purely oscillatory (tidal) flow in two dimensions, we suggest setting
\begin{align*}
    F_{\mathrm{2d}}&=\frac{H}{H-h_0} F_{\text{SAH2d}}, &\text{if}\ h_0\leq H/2,\\
    F_{\mathrm{2d}}&=2\: F_{\text{SAH2d}}, &\text{if}\ h_0> H/2,
\end{align*}
noting that this parametrisation may overestimate the stress when $h_0\rightarrow H$.

In three dimensions, we then similarly suggest setting
\begin{align*}
    F_{\mathrm{3d}}&=\frac{H}{H-h_0} F_{\text{SAH3d}}, &\text{if}\ h_0\leq H/2,\\
    F_{\mathrm{3d}}&=2\: F_{\text{SAH3d}}, &\text{if}\ h_0> H/2,
\end{align*}
for large hill widths $W$, where $F_{\text{SAH3d}}$ is as in \eqref{SAH3d}. From our testing, $W=10000\ \mathrm{m}$ was a sufficiently large width to notice this effect (with $h_0=100\ \mathrm{m}$). For more modest values of $W$, we instead suggest setting $F_{3d}=F_{\text{SAH3d}}$.

\subsection{Suggested steady flow parametrisations} For a purely steady flow, we found that the parametrisations due to \citet{bell1975topographically} and \citet{klymak2010high} had mostly accurate scalings. However, to better match the simulation data we suggest introducing additional parameters that account for non-linear effects and need to be suitably chosen.

For a purely steady flow in two dimensions, we suggest setting
\begin{align}
    F_{\text{2d}}&=\frac{1}{A_1-B_1 \, Fr}F_{\text{Bell2d}},& \text{if}\ Fr\lesssim 1\label{2dssug1},\\
    F_{\text{2d}}&=C_1Nh_0^2U,&\text{if}\ Fr\gg 1\label{2dssug2},
\end{align}
where $A_1$, $B_1$ and $C_1$ are tunable constants, $Fr=Nh_0/U$, and $F_{\text{Bell2d}}$ is as defined in \eqref{bell-full2d}. Our data gave the fitted values $A_1=0.71$, $B_1=0.46$ and $C_1=1.4$. Note that one can use the Taylor expansion of~\eqref{2dssug1} for small Froude numbers ($Fr \ll 1$) to obtain an expression with polynomial dependence on $Fr$. The extreme cases of low and high Froude numbers covers many possible regimes in the real world, for example abyssal hills for low $Fr$ and seamounts for large $Fr$. However, we did not analyse the transitional case of medium Froude numbers ($Fr\approx 1-10$) in detail. For this, we refer the reader to the recent work of \citet{klymak2021parameterizing} where a hybrid parametrisation is tested to account for this additional regime.

In three-dimensions, we then suggest setting
\begin{align}
    F_{\text{3d}}&=\frac{1}{A_2-B_2|fW/U|}F_{\text{Bell3d}},\ \text{if}\ Fr\lesssim 1,\ \text{and}\ |fW/U|\lesssim 1,\label{3dssug1}\\
    F_{\text{3d}}&=C_2h_0WU^2,\quad\text{if}\ Fr\gg 1\label{3dssug2},
\end{align}
where similarly $A_2$, $B_2$ and $C_2$ are tunable constants, and $F_{\text{Bell3d}}$ is as defined in \eqref{bell-full3d}. Our data gave the fitted values $A_2=1.34$, $B_2=0.88$ and $C_2=1$. Similar to \eqref{2dssug1}, one could also change \eqref{3dssug1} to have a polynomial dependence on $|fW/U|$ if desired. Such an alternate form of \eqref{3dssug1} could be useful in the regime $A_2-B_2|fW/U|\approx 0$, in which \eqref{3dssug1} is unphysical.

\subsection{Suggested mixed flow parametrisations}
For a mixed flow
\begin{equation}U(t)=U_{\text{mean}}+U_{\text{tidal}}\cos(\omega t),
\end{equation}
with associated stress (parallel to $U(t)$)
\begin{align*}
    F_{2d}&=F_m+F_p\cos(\omega t+\phi),\qquad\text{(two dimensions)}\\
    F_{3d}&=F_m+F_p\cos(\omega t+\phi),\qquad\text{(three dimensions)}
\end{align*}
the naive parametrisation
\begin{equation}\label{naivesug}
    F_m=\widetilde{F}_m,\qquad F_p=\widetilde{F}_p
\end{equation}
proved to be accurate for a wide range of parameters. Here, as in Sections~4\ref{mixedtworesults} and~4\ref{mixedthreeresults}, $\widetilde{F}_m$ and $\widetilde{F}_p$ are the values of $F_m$ and $F_p$ if $U_{\text{tidal}}$ or $U_{\text{mean}}$ are set to be zero respectively. In two dimensions, the main regimes where \eqref{naivesug} disagreed with, and notably underestimated, our simulation output were those near the critical latitude $|f|=\omega$, those with small values of $U_{\text{mean}}$ (for $F_m$), and those with small values for $W$ (for $F_p$). Then, in three dimensions, the naive parametrisation again underestimated the simulation data near the critical latitude $|f|=\omega$ (for $F_p$), and also for large hill heights $h_0$ (for $F_m$). For this last point, we suggest setting
\begin{equation}\label{largefmixedsug}
F_m=\widetilde{F}_m \left(1+\frac{U_{\text{tidal}}^2}{2U_{\text{mean}}^2} \right),
\end{equation}
in three dimensions whenever the Froude number ${Fr=Nh_0/U_{\text{mean}}}$ is large. The formula \eqref{largefmixedsug} is derived from the quadratic drag law \eqref{3dssug2} we observed in the steady flow case for $Fr\gg 1$. However, as discussed in Section~\ref{resultssect}f, it would be preferable to do more tests and analysis to evaluate the accuracy of \eqref{largefmixedsug}, and to better understand the transition between the low and high Froude number regimes.

In the bottom-trapped regime ($|f|>\omega$), the phase $\phi$ also notably disagreed with that predicted by the \citet{shakespeare2020drag} theory for tidal flows, but still satisfied $\phi\approx 0$ for $|f|\ll\omega$.

\section{Conclusion}
We have evaluated common stress parametrisations for tidal, steady and mixed flows over rough topography. In the case where the topography is an isolated Gaussian hill in two or three dimensions, we have directly compared these parametrisations with idealised numerical simulations. This helped clarify the cases for which the parametrisations and simulations agreed and disagreed. We found that the most notable regimes where the parametrisations and simulations disagreed were (i) when the topographic height $h_0$ was large, and (ii), when the magnitude of the Coriolis parameter $|f|$ was close to the tidal frequency $\omega$ (for tidal and mixed flows). This was not entirely unexpected, as the existing parametrisations were often designed for small hill heights $h_0$ or in the case where~$|f|\ll\omega$.

We also suggested simple adjustments or new parametrisations in the cases where the existing theory falls short. In most cases, only simple adjustments to the standard theory was required, highlighting the relative success of the underlying linear theory, even in regimes considered to be strongly non-linear. It is hoped that these updated parametrisations may be used to more accurately model stress in regimes for which there is currently limited theory available. However, before applying our work to large-scale ocean models, there are some notable limitations that need further consideration.

Firstly, one would need to test how well our scalings and parametrisations hold with more complicated topography, beyond the isolated hill scenario that we studied. For example, as we mentioned in the text, \citet{klymak2021parameterizing} found that for steady flows in three-dimensions, the stress varies differently with the buoyancy frequency $N$ depending on whether one has an isolated obstacle, or multiple randomly distributed obstacles. Our assumption that $N$ is uniform should also be relaxed, and as discussed by \citet{luschow2024sensitivity}, the choice of $N$ that one uses in a parametrisation can significantly affect the accuracy of the model.

Secondly, there are some specific parameter regimes where more analysis is needed. In particular, more investigation is needed of  steady flow regimes where the Froude number $Fr$ has an intermediate value, rather than just the limits $Fr\lesssim1$ and $Fr\gg 1$ considered here.

Lastly, a critical step in implementing any stress parameterisation is determining where in the vertical the stress should be applied to the flow, which was not considered here. However, previous studies have provided guidance on this point. For mean stress, the propagating wave component is applied where the lee waves break and dissipate \citep[e.g.,][]{booker1967critical,andrews1976planetary,xie2015generalised}, while the non-propagating part acts directly adjoining the hill \citep[e.g.,][]{klymak2010high,klymak2018nonpropagating,klymak2021parameterizing}.  By contrast, for periodic stress, the force acts in the diffusive boundary layer adjoining the topography as shown in \citet{shakespeare2021impact}. However, no previous studies have considered the mixed flow case, in which this picture may differ, and this is clearly an important avenue for further investigation.

Despite these limitations, the present work provides an important step towards a more comprehensive and robust parametrisation of topographic stresses in ocean models.

\acknowledgments
We thank Nicolas Grisouard and the two anonymous reviewers for their constructive comments. D.R.J.~would also like to thank Geoff Stanley and John Scinocca for sharing their thoughts on an early draft of this paper at the CCCma in Victoria, Canada during June 2024. N.C.C.~acknowledges support from the Australian Research Council under DECRA Fellowship DE210100749, Discovery Project~DP240101274, and the Center of Excellence for the Weather of the 21st Century CE230100012.

%
%
\datastatement
Scripts that reproduce the simulations and all the data used in this paper are available at the repository \url{github.com/DJmath1729/Ocean-hill-stress} and also at Zenodo \citep{johnston_2025_17274188}. Oceananigans is available at \url{github.com/CliMA/Oceananigans.jl}; version 0.82 was used for this paper.

\end{document}